\documentclass[twocolumn,pra,preprintnumbers,showpacs,superscriptaddress,10pt]{revtex4-1}
\usepackage{hyperref}
\usepackage{verbatim}
\usepackage{amsmath}
\usepackage{latexsym}
\usepackage{natbib}
\usepackage{amsmath}
\usepackage{amssymb}
\usepackage{amsthm}
\usepackage{graphicx}
\usepackage{pict2e} 
\usepackage{rotating} 
\usepackage{bm}
\usepackage{color}

\newcommand{\figref}[1]{Figure~\ref{#1}}

\newcommand{\id}{\mathbb{I}}
\newcommand{\tr}{\mathop{\mathrm{tr}}\nolimits}

\newcommand{\ra}{\rightarrow}
\newcommand{\hil}{\mathcal{H}}
\newcommand{\kil}{\mathcal{K}}
\newcommand{\group}{\mathcal{G}}
\newcommand{\nn}{\nonumber}
\newcommand{\eps}{\varepsilon}
\newcommand{\half}{\frac{1}{2}}
\newcommand{\mdag}{^{\dagger}}
\newcommand{\m}[1]{\mathrm{#1}}
\newcommand{\cL}{\ensuremath{\mathcal{L}}}
\DeclareMathOperator{\supp}{supp}

\newcommand{\ket}[1]{|#1\rangle}

\usepackage{amsfonts}

\def\Complex{\mathbb{C}}

\def\id{\mathbb{I}}

\begin{document}

\title{Self-correcting quantum memory with a boundary}
\author{Adrian Hutter}\email{adrian.hutter@unibas.ch}
\affiliation{Department of Physics, University of Basel, Klingelbergstrasse 82, CH-4056 Basel, Switzerland}
\author{James R.\ Wootton}
\affiliation{Department of Physics, University of Basel, Klingelbergstrasse 82, CH-4056 Basel, Switzerland}
\author{Beat R\"othlisberger}
\affiliation{Department of Physics, University of Basel, Klingelbergstrasse 82, CH-4056 Basel, Switzerland}
\author{Daniel Loss}
\affiliation{Department of Physics, University of Basel, Klingelbergstrasse 82, CH-4056 Basel, Switzerland}

\date{\today}

\begin{abstract}
We study the two-dimensional toric code Hamiltonian with effective long-range interactions between its anyonic excitations induced by coupling the toric code to external fields.
It has been shown that such interactions allow to increase the lifetime of the stored quantum information arbitrarily by making $L$, the linear size of the memory, larger [Phys.\ Rev.\ A \textbf{82} 022305 (2010)].
We show that for these systems the choice of boundary conditions (open boundaries as opposed to periodic boundary conditions) is not a mere technicality; 
the influence of anyons produced at the boundaries becomes in fact \emph{dominant} for large enough $L$.
This influence can be both beneficial or detrimental.
In particular, we study an effective Hamiltonian proposed in [Phys.\ Rev.\ B \textbf{83} 115415 (2011)] that describes repulsion between anyons and anyon holes.
For this system, we find a lifetime of the stored quantum information that grows exponentially in $L^2$ for both periodic and open boundary conditions, though the exponent in the latter case is found to be less favourable.
However, $L$ is upper-bounded through the breakdown of the perturbative treatment of the underlying Hamiltonian.
\end{abstract}


\maketitle

\section{Introduction}

An important open problem in quantum information concerns the feasibility of a \emph{self-correcting quantum memory}. 
Finding a system that protects a quantum state from decoherence induced by a thermal bath, without the need for active monitoring and error-correction, proves much more difficult than in the classical case.
On the most basic level, this is due to the fact that a classical bit only needs protection against logical $X$ operations while a qubit needs protection against a logical $X$ and $Z$.
If a state is stored in a many-qubit system, a desirable feature is \emph{topological protection} of the stored (qu-)bit: 
we want a logical error $X$ (or $Z$ in the quantum case) to necessitate a number of single-qubit errors $\sigma_x$ (or $\sigma_z$) that scales with $L$, the linear size of the memory.
The simplest model that energetically penalizes $\sigma_x$ errors and offers topological protection of a stored classical bit is the 1D ferromagnetic Ising model. 
The simplest model that penalizes $\sigma_x$ and $\sigma_z$ errors and topologically protects a qubit is given by Kitaev's 2D toric code Hamiltonian \cite{Kitaev03}.
In fact, the latter can be mapped exactly to two independent copies of the former \cite{Nussinov07,Alicki09}.
Unfortunately, both of these systems are not thermally stable. 
Once a topological defect (a pair of domain walls in the 1D Ising model or a pair of anyons in the 2D toric code) has been created, it can spread and lead to a logical error without any further energy cost.
The lifetime of a qubit stored in the degenerate ground states of the 2D toric code is thus for any finite temperature upper-bounded by a constant independent of $L$ \cite{Nussinov08,Alicki09,Chesi10}.
The 2D toric code can therefore not serve as a `quantum hard drive'.
These difficulties can be overcome if the dimensionality of the systems is increased. 
In the 2D Ising model and the 4D toric code, any sequence of single-qubit Pauli operators that leads to a logical error has to surpass an energy barrier whose size scales with $L$. 
Since the number of error paths connecting two distinct ground states is exponentially large in $L$, these systems are thermally stable below some critical temperature $T_c$, meaning that the lifetime of the stored information grows exponentially with $L$ \cite{Dennis02,Alicki09,Alicki10}.

Whether a similar degree of protection for a quantum state can be achieved in less than four dimensions is not clear.
One can show that for \emph{every} 2D local stabilizer Hamiltonian the height of the energy barrier separating orthogonal states stored in a degenerate ground state is upper bounded by a constant independent of $L$
\cite{Kay08,Bravyi09,Haah10}, ruling out the possibility of using such systems for the fault-tolerant storage of quantum information by self-correction.
In principle, these no-go results leave two ways out: Either one abandons the locality of the terms in the Hamiltonian or one goes to dimension 3.
Indeed, both of these routes have been followed in the recent literature.
While the 3D toric code is not thermally stable, Haah showed in a recent breakthrough the existence of 3D Hamiltonians with local interactions that have no string-like logical operators \cite{Haah11a}.
Unlike anyons in the 2D toric code, defects cannot move further than a certain constant distance away without creating other defects. 
This property implies a logarithmically growing energy barrier between orthogonal ground states, leading one to expect a lifetime that grows polynomially with $L$ \cite{Haah11b}.
However, the best known lower bound on the memory lifetime of Haah's Hamiltonian is upper-bounded by a constant independent of $L$ \cite{Haah11c} and further improvement is not expected \cite{communication_comment}.
Furthermore, a 3D architecture may lead to practical difficulties when accessing the physical qubits for syndrome measurement and error correction.

We therefore believe that the most promising route to follow in search for a realistic proposal for a quantum memory is to start from the 2D toric code Hamiltonian and add terms to it
that
\begin{itemize}
 \item can be physically motivated, and
 \item lead to a memory lifetime that becomes arbitrarily large as $L\ra\infty$\ .
\end{itemize}
Long-range repulsive interactions ($1/r^\alpha$-potential with $0\leq\alpha<2$) between the anyons lead to a logarithmically-growing self-consistent mean field gap for anyon creation, yielding a polynomially increasing lifetime \cite{Chesi10}, see Sec.~\ref{sec:longrange} below.
So rather than trying to find a Hamiltonian with a macroscopic energy barrier between orthogonal ground states, this approach seeks to suppress the anyon creation rate.
The toric code Hamiltonian (involving local \emph{four}-qubit couplings) with non-interacting anyons can be obtained as an effective Hamiltonian of the Kitaev honeycomb model, which involves nearest-neighbor \emph{two}-qubit Ising couplings \cite{Kitaev06}. 
How an $\alpha = 0$ interaction between the anyons can be obtained through such a honeycomb model coupled to electromagnetic modes has been studied in detail in \cite{Pedrocchi11}, see Sec.~\ref{sec:honeycomb} below.

In an alternative approach it was shown that coupling the toric code to a bosonic field leads to an effective gravitational potential between the anyonic defects \cite{Hamma09}. 
Below some critical temperature, all anyons coalesce to a single point. However, the time the system needs to approach this metastable  state and how to best perform error correction in this system have not been investigated so far.

Self-correcting quantum memories are usually discussed with periodic boundary conditions, giving the toric code its name.
This way, the complications that arise with the possibility of creating unpaired topological defects at the boundaries can be avoided.
One expects that the influence of the boundaries becomes negligible if $L$ becomes large enough, which is certainly correct for Hamiltonians with local interactions.
However, here we study memories with long-range interactions between the anyons as proposed in \cite{Chesi10,Pedrocchi11} and show that for these systems the influence of the boundary becomes in fact \emph{dominant} for large enough $L$. 
It can be beneficial and detrimental. Specifically, unpaired anyons from the boundary lead to an effective bias for anyons from the bulk to move to the closest boundary, thus prolonging the time until error correction becomes ambiguous (see Sec.~\ref{sec:longrange}). 
On the other hand, the ability to create unpaired anyons at the boundaries halves the energetic gap above the anyonic vacuum. 
This becomes especially relevant if this gap is so strong that the anyonic system is basically restricted to its ground state and first excited state (see Sec.~\ref{sec:honeycomb}).

The paper is structured as follows. In Section \ref{sec:correction} we discuss how error correction can be performed in the planar code (a toric code with open boundaries) in contact with a thermal environment.
In Section \ref{sec:longrange} we study the influence of the boundaries for a Hamiltonian with spatially constant repulsion between the anyons, while in Section \ref{sec:honeycomb} an effective Hamiltonian that describes repulsion between anyons and anyon holes is investigated.

\section{Error correction in the planar code}\label{sec:correction}

\subsection{The planar code}

\begin{figure}
	\includegraphics[width=0.6\textwidth]{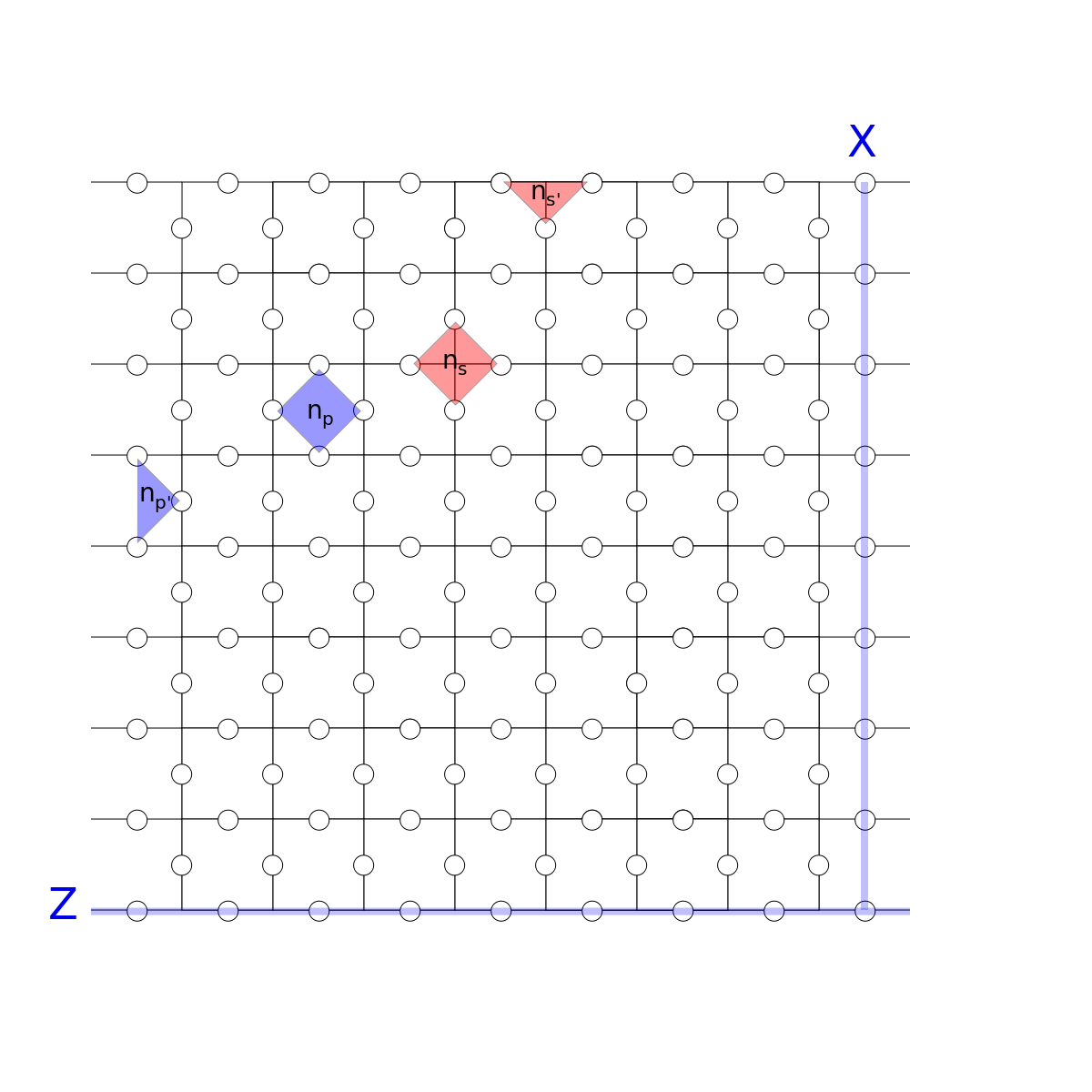}
	\caption{A planar code of size $L=8$. Depicted are a four-qubit plaquette operator $n_p$, a three-qubit plaquette operator $n_{p'}$, a four-qubit star operator $n_s$, and a three-qubit star operator $n_{s'}$. 
                 The logical operator $X$ ($Z$) is given by any chain of Pauli operators $\sigma_x$ ($\sigma_z$) connecting the top and bottom (left and right) boundary.}
	\label{fig:stabilizers}
\end{figure}

A self-correcting quantum memory is supposed to protect a quantum state from a thermal environment by means of its internal dynamics and without need for active error-correction.
A single error correction step may be performed before the stored state is read out. We shall use here a version of the toric code first introduced in \cite{Bravyi98}, which, contrary to what the name suggests, is not periodic but does have a boundary.
We will refer to this as the \emph{planar code}. Consider a grid with quadratic cells and physical qubits placed on the edges, as depicted in \figref{fig:stabilizers}.
We call the four qubits around one unit cell a `plaquette' and the four qubits around a vertex a `star'. 
We define plaquette operators $A_p=(\sigma_z)^{\otimes4}$, where the tensor product runs over the four qubits around some plaquette $p$ and star operators $B_s=(\sigma_x)^{\otimes4}$, where the tensor product runs over the four qubits around some vertex $s$. 
Plaquette operators on the left and right boundary and star operators on the top an bottom boundary are tensor products of three Pauli operators only.
All of these operators are commuting since they overlap at zero or two qubits.
Let the space $\kil_0\subset\hil=(\Complex^2)^{\otimes N}$ ($N$ is the total number of qubits) be defined as the space stabilized by all plaquette and star operators. 
That is, $\kil_0$ is the space of all states $\ket{\psi}$ such that for each three- or four-qubit plaquette or star operator $A_p$ or $B_s$ we have $A_p\ket{\psi}=\ket{\psi}$ and $B_s\ket{\psi}=\ket{\psi}$.
Since all the plaquette and star operators are independent (unlike in the toric code where the product of all plaquette and star stabilizer operators is the identity), one easily verifies that $\dim\kil_0=2$, independent of the height and width of the grid \cite{eliminates_comment}, such that one logical qubit can be stored in this space.
States in $\kil_0$ are \emph{topologically protected}. They cannot be distinguished by any local observable and not be evolved into each other by any local unitary.

We then define a Hamiltonian that imposes an energy penalty for the violation of every stabilizer condition. 
Let $n_p = (1-(\sigma_z)^{\otimes4})/2$, $n_{p'} = (1-(\sigma_z)^{\otimes3})/2$, $n_s = (1-(\sigma_x)^{\otimes4})/2$, and $n_{s'} = (1-(\sigma_x)^{\otimes3})/2$. 
The tensor products run over the qubits depicted in \figref{fig:stabilizers}.
These operators have eigenvalue $0$ for states that satisfy the corresponding stabilizer conditions and eigenvalue $1$ for states that violate it.
For some state $\ket{\psi}\in\hil$ we say that an anyon is present at plaquette $p$ (vertex $s$) if $n_p\ket{\psi}=\ket{\psi}$ ($n_s\ket{\psi}=\ket{\psi}$) and that no anyon is present if $n_p\ket{\psi}=0$ ($n_s\ket{\psi}=0$),
i.e.\ we interpret stabilizer violations as the presence of anyons.
We then use the well-known toric code Hamiltonian
\begin{align}\label{eq:Kitaev}
 H_{\m{Kitaev}} = \Delta\cdot\left(\sum_pn_p+\sum_sn_s\right)\ ,
\end{align}
which simply counts the total number of anyons. The code subspace $\kil_0$, which is the degenerate ground state of this Hamiltonian, corresponds to the anyonic vacuum. 
This Hamiltonian is stable against weak local perturbations in the sense that the lifting of the ground state degeneracy through such a perturbation is exponentially small in $L$ \cite{Kitaev03}.

Starting from the anyonic vacuum, if a qubit suffers a spin-flip error $\sigma_x$ (phase-flip error $\sigma_z$), two plaquette (star) anyons are created on the two adjacent plaquettes (vertices). 
Once an anyon exists, it can move on the surface without any further energy cost, whereby creating further spin- or phase-flip errors along its path. Two anyons of the same kind can fuse to the vacuum.
On the two horizontal (vertical) boundaries of the grid in Figure~\ref{fig:stabilizers} a single plaquette (star) anyon can be created. Similarly, the boundaries can absorb single anyons.
If two anyons are jointly created from the vacuum, move around and then fuse to the vacuum, the produced error path can be expressed as a product of stabilizers and therefore acts trivially on the code subspace $\kil_0$. 
The same holds if a single anyon is created on a boundary and then is absorbed by the same boundary again. The only possibility to act non-trivially on the state stored in $\kil_0$ is if an error path connects the two opposite boundaries.
We therefore define the logical operators $X = \bigotimes\sigma_x$ and $Z = \bigotimes\sigma_z$ acting on all qubits along the paths depicted in \figref{fig:stabilizers}. 
Since all such products that connect the two opposite boundaries are identical up to multiplication with stabilizers, the precise form of the path does not matter.
These operators commute with all the stabilizers (i.e.\ they are elements of the centralizer of the stabilizer group) though are not products of stabilizers. They allow thus to act non-trivially on the state stored in $\kil_0$ without leaving any smoking guns in the form of anyons.
They satisfy $X^2 = Z^2 = \id$ and $XZ = -ZX$ and may therefore be seen as Pauli operators acting on the encoded qubit.

At the read-out step, all the operators $n_s$ and $n_p$ are measured and the presence of anyons (the \emph{syndrome}) is detected. 
The goal is then to annihilate all anyons (either by fusing them to the vacuum or by moving them to a boundary that can absorb them) and thereby to undo the errors caused by the diffusion of the anyons.
More precisely, the goal is that the total unitary formed by the natural anyon dynamics (creation, diffusion, and annihilation) plus the error correction procedure is equal to a product of stabilizers and thus acts trivially on the code space.
Even more precisely, the dynamics induced by the thermal environment are in fact a probabilistic mixture of different unitary evolutions, this will be discussed in more detail later on.

\subsection{Error correction}

We will study planar codes of different sizes $L$, by which we mean that the number of plaquette operators from top to bottom as well as the number of star operators from left to right is given by $L$.
Consequently, there are $L-1$ four-qubit plaquettes and $2$ three-qubit plaquettes from left to right and thus a total of $L\cdot(L+1)$ plaquettes.
The total number of qubits is then given by $N= 2L^2+2L+1$.
It is well-known that in the limit of large $L$ error correction in the toric code is unambiguously possible if less than $11\%$ of the physical qubits are subject to uncorrelated bit- and phase-flip errors \cite{Dennis02}.
However, qubit errors induced by a thermal environment \emph{are} correlated due to the diffusion of the anyons. Error correction therefore typically becomes impossible if a few percent of the qubits have suffered errors \cite{Chesi10}.

The error correction step consists of three substeps. In a first step, the anyon configuration or syndrome is determined by measuring all the operators $n_s$ and $n_p$.
This means that we project to the subspaces $\kil_i$ with a given anyon configuration. Since there are $L\cdot(L+1)$ potential locations for anyons of each kind, there are $2^{2L\cdot(L+1)}$ such spaces, each of which has dimension $2$ 
(they can be obtained from $\kil_0$ by applying single-qubit errors), so
\begin{align}
 \hil = \bigoplus_{i=0}^{2^{2L\cdot(L+1)}-1}\kil_i\ .
\end{align}
Indeed, one verifies that $2\cdot2^{2L\cdot(L+1)}$ equals $\dim\hil = 2^N = 2^{2L^2+2L+1}$. 
In a second step, a classical computation is performed on the error syndrome whose output tells how to best annihilate the anyons by fusing them with each other or moving them to a boundary of the type that can absorb them, 
which is then done in a third step.

In a more formal language, let $\rho_0$ denote the initial state stored in the memory with $\supp\rho_0\subseteq\kil_0$. The influence of errors on this state is then captured by a quantum channel (CPTPM) $\Phi_{\m{err}}$.
The goal of the classical computation is then, given knowledge about the error model $\Phi_{\m{err}}$ and the syndrome (the space $\kil_i$, that is), to find a sequence of single-qubit Pauli operators $U_i$ which corrects the errors.
In other words, we want $U_i$  to map $\kil_i$ to $\kil_0$.
Now let $P_i$ denote the projector onto $\kil_i$.
Formally, we can write the error-correction procedure performed on the corrupted state $\rho_t = \Phi_{\m{err}}^t(\rho_0)$ as $\Phi_{\m{corr}}(\rho_t) = \sum_i U_iP_i\rho_tP_iU_i\mdag$.
At the end of the day, we want the error
\begin{align}
 \delta(t) := \left\|\rho_0 - \left(\Phi_{\m{corr}}\circ\Phi_{\m{err}}^t\right)(\rho_0)\right\|_1
\end{align}
to be as small as possible for any given encoded state $\rho_0$.
In the corrected state $\left(\Phi_{\m{corr}}\circ\Phi_{\m{err}}^t\right)(\rho_0)$ no anyons are left. We therefore either have successfully corrected all errors, performed a logical $X$, a logical $Z$ or both, thus
\begin{align}
 &\left(\Phi_{\m{corr}}\circ\Phi_{\m{err}}^t\right)(\rho_0) \nn\\&\quad = (1-p_X)(1-p_Z)\cdot\rho_0 + p_X(1-p_Z)\cdot X\rho_0X \nn\\&\qquad+ (1-p_X)p_Z\cdot Z\rho_0Z + p_Xp_Z\cdot XZ\rho_0ZX
\end{align}
(assuming that the error correction procedure treats plaquette- and star-anyons independently).
We therefore have
\begin{align}\label{eq:totalError}
 \delta(t) &\leq 2\cdot\left(p_X(1-p_Z)+(1-p_X)p_Z+p_Xp_Z\right) \nn\\&\leq 2(p_X+p_Z)\ .
\end{align}

In order to obtain simple scalar functions that characterize the decay of the stored quantum information, we study the autocorrelation functions
\begin{align}
 &C^X_{\m{corr}}(t) := 2^{-N}\tr\left[X\cdot\left(\Phi_{\m{corr}}\circ\Phi_{\m{err}}^t\right)\mdag(X)\right]\\
 &C^Z_{\m{corr}}(t) := 2^{-N}\tr\left[Z\cdot\left(\Phi_{\m{corr}}\circ\Phi_{\m{err}}^t\right)\mdag(Z)\right]\ .\label{eq:CZcorr}
\end{align}
The prefactor is such that $C^X_{\m{corr}}(0) = C^Z_{\m{corr}}(0) = 1$, assuming that no operation is performed on the stored information if no anyons are measured.

This assumption is in fact less trivial than it may seem. Performing a logical operation in the error correction step is beneficial if an odd number of logical operators have been performed by the bath. 
Assume that the bath induces logical errors that leave no anyons with rate $r$.
Then, the probability that no logical error has been performed is in fact small for times $t\gg r^{-1}$.
The probability that after time $t$ $k$ logical errors have been performed is given by the Poisson distribution,
\begin{align}
 P(k, rt) = \frac{(rt)^ke^{-rt}}{k!}\ .
\end{align}
The Poisson distribution is peaked around $rt$, which may be an odd integer. However, the probability
\begin{align}\label{eq:PoissonEven}
\sum_{k\, \m{even}} 
P(k, rt) = \half\left( 1 + e^{-2rt} \right)
\end{align} 
of an even number of errors is greater than $\half$ for any $rt$, such that the optimal strategy is, indeed, not to do anything if no anyon is detected.

$C^X_{\m{corr}}(t)$ is $1$ if after error correction at time $t$ no logical $Z$-operator has been applied and $-1$ if one has been applied (i.e.\ an odd number of $\sigma_z$-operators has been applied to any line connecting the left and right boundary).
Therefore, $C^X_{\m{corr}}(t) = 1-2p_Z$ and analogously $C^Z_{\m{corr}}(t) = 1-2p_X$.
In conclusion we have
\begin{align}\label{eq:CcorrP}
 &\left\|\rho_0 - \left(\Phi_{\m{corr}}\circ\Phi_{\m{err}}^t\right)(\rho_0)\right\|_1
 \nn\\&\quad\leq \left(1-C^X_{\m{corr}}(t)\right) + \left(1-C^Z_{\m{corr}}(t)\right)\ .
\end{align}
We define the lifetime $\tau(\eps)$ of the memory as the maximal time such that $\min\left\{C^X_{\m{corr}}(t) , C^Z_{\m{corr}}(t)\right\} \geq 1-\eps$ for all $t\leq\tau(\eps)$, implying that $\delta(t)\leq2\eps$ for $t\leq\tau(\eps)$.

The total evolution $\Phi_{\m{corr}}\circ\Phi_{\m{err}}^t$ is a statistical mixture of different unitary evolutions.
In the numerical simulations, we will in each run follow a definite unitary evolution, such that $C^Z_{\m{corr}}(t)$ is at any time given by $\pm1$.
The probability of a certain unitary is thereby determined by the error model $\Phi_{\m{err}}$. Sampling over a large number of runs, we obtain a smooth function $C^Z_{\m{corr}}(t)$.

We say that two sequences of single-qubit Pauli operators are equivalent if they are identical up to multiplication with stabilizers.
For every given anyon configuration, there are four equivalence classes of errors that produce it from the vacuum. These equivalence classes can be mapped onto each other by application of the logical operators $I$, $X$, $Z$, $XZ$.
Given a syndrome, the goal is to guess the most likely equivalence class of errors that has produced it, which one allows to remove the anyons without disturbing the stored quantum information.
Calculating the probabilities of the four equivalence classes is numerically too costly to be performed with current technology.
We thus make the simplifying assumption that the most likely error path that has led to the given syndrome is an element of the most likely equivalence class.
This may not be true for every possible anyon configuration but seems a reasonable approximation. Applying stabilizers to the most likely error path will produce further error paths with identical or slightly lower probabilities that are elements of the same equivalence class.
Rather than finding the error path with maximal probability, we may equivalently find the error path with \emph{minimal weight}, if we define the weight to be the negative logarithm of the probability that a certain error chain has occurred.
Taking the negative logarithm ensures that the weight is additive for independent error chains. This is known as the Shannon information content of an event in classical information theory \cite{MacKay02} and up to a constant factor the only function having the additivity property.
 
To illustrate this, let us consider a concrete simple error model. We assume that each physical qubit suffers a spin-flip error with probability $p_x$ and a phase-flip error with probability $p_z$,
\begin{align}
 &\Phi_{\m{err}} = \Phi_{\m{single-qubit}}^{\otimes N} \quad\m{,}\nn\\
 &\Phi_{\m{single-qubit}}(\omega) \nn\\&\quad = (1-p_x)(1-p_z)\cdot\omega + p_x(1-p_z)\cdot\sigma_x\omega\sigma_x \nn\\&\qquad\ + (1-p_x)p_z\cdot\sigma_z\omega\sigma_z + p_xp_z\cdot\sigma_y\omega\sigma_y\ .
\end{align}
The weight of an error chain involving $\ell_x$ spin-flips and $\ell_z$ phase-flips is then 
\begin{align}\label{eq:indep_weight}
 \ell_x\cdot\ln\frac{1-p_x}{p_x} + \ell_z\cdot\ln\frac{1-p_z}{p_z} + \m{const}\ ,
\end{align}
allowing us to minimize (for $p_x , p_z < \half$) $\ell_x$ and $\ell_x$ independently.
The number of single-qubit Pauli operators necessary to connect two anyons with each other is given by the so-called `Manhattan distance' of the anyons, i.e.\ the sum of the horizontal and the vertical coordinate difference of two anyons.
Similarly, the weight of a chain connecting an anyon to a boundary is given by the horizontal or vertical distance.
In detail, our minimal weight matching of $n$ anyons (of one kind) then works as follows.
\begin{enumerate}
 \item Perform a Delaunay triangulation on the set of anyon coordinates, thereby restricting the full graph of $\frac{n(n-1)}{2}$ edges between anyons to $O(n)$ edges. Calculate the Manhattan weights of all edges in the restricted graph.
 \item For every anyon, add a `virtual' partner on the closer boundary able to absorb it and add an edge to the graph with weight given by the distance to the boundary.
 \item Connect all virtual anyons with zero-weight edges.
 \item Perform a minimum-weight matching of the graph obtained this way.
\end{enumerate}
Points 1.\ and 4.\ are identical to the methods used in \cite{Chesi10,Beat12}. 
We perform the Delaunay triangulation using the library \texttt{Triangle} \cite{Shewchuk96} while for the minimal-weight perfect matching we employ the library \texttt{Blossom V} \cite{Kolmogorov09} implementing the `blossom' algorithm due to Edmond's \cite{Edmonds65}.
The numerical cost of this procedure is strongly dominated by the last step.
Adding the virtual anyons ensures that each real anyon can be connected to the closest edge able to absorb it and that there is always an even number of points in the graph entering the perfect matching algorithm.
Giving the edges between virtual anyons weight zero ensures that those virtual anyons that are not connected to a real one can be removed at no cost.

\begin{figure}
  \setlength{\unitlength}{0.5\textwidth}
    \begin{picture}(0.8,0.8)	
	\put(0.01,0.1){\includegraphics[width=0.40\textwidth]{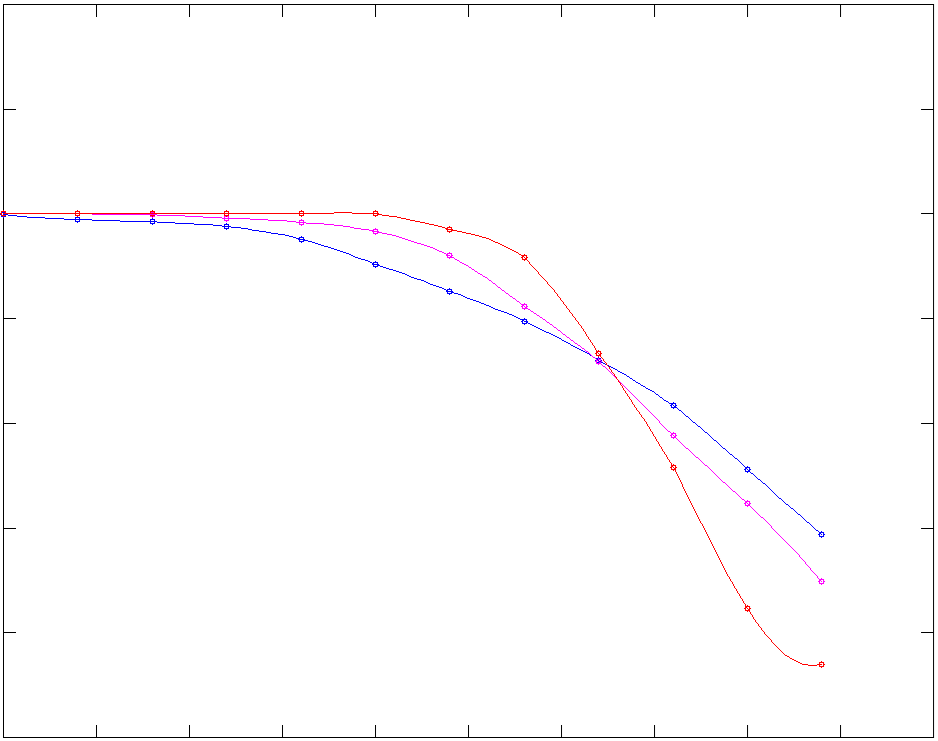}}
	\put(-0.09,0.40){\begin{sideways}$C^X_{\m{corr}}$\end{sideways}}
	\put(0.36,0.03){$p_z$}
	
	\put(-0.01,0.07){\scriptsize $0.07$}
	\put(0.06,0.07){\scriptsize $0.075$}
	\put(0.15,0.07){\scriptsize $0.08$}
	\put(0.22,0.07){\scriptsize $0.085$}
	\put(0.31,0.07){\scriptsize $0.09$}
	\put(0.38,0.07){\scriptsize $0.095$}
	\put(0.47,0.07){\scriptsize $0.10$}
	\put(0.54,0.07){\scriptsize $0.105$}
	\put(0.63,0.07){\scriptsize $0.11$}
	\put(0.70,0.07){\scriptsize $0.115$}
	\put(0.79,0.07){\scriptsize $0.12$}

	\put(-0.04,0.72){\scriptsize $1.4$}
	\put(-0.04,0.63){\scriptsize $1.2$}
	\put(-0.04,0.54){\scriptsize $1.0$}
	\put(-0.04,0.45){\scriptsize $0.8$}
	\put(-0.04,0.365){\scriptsize $0.6$}
	\put(-0.04,0.275){\scriptsize $0.4$}
	\put(-0.04,0.185){\scriptsize $0.2$}
	\put(-0.04,0.100){\scriptsize $0.0$}

    \end{picture}
  \caption{The plot shows the autocorrelation function $C^X_{\m{corr}}$ as a function of the fraction of physical spins $p_z$ that have suffered spin-flip errors. 
		Circles give the numerical results, the lines are guides to the eye.
		The curves correspond to different lattice sizes $L = 32$ (blue), $L = 64$ (magenta) and $L = 128$ (red).
		  We anticipate that in the limit $L\ra\infty$ error correction is unambiguously possible for any $p_z < p_c$, with $p_c \gtrsim 0.102$.}
  \label{fig:decay}
\end{figure}
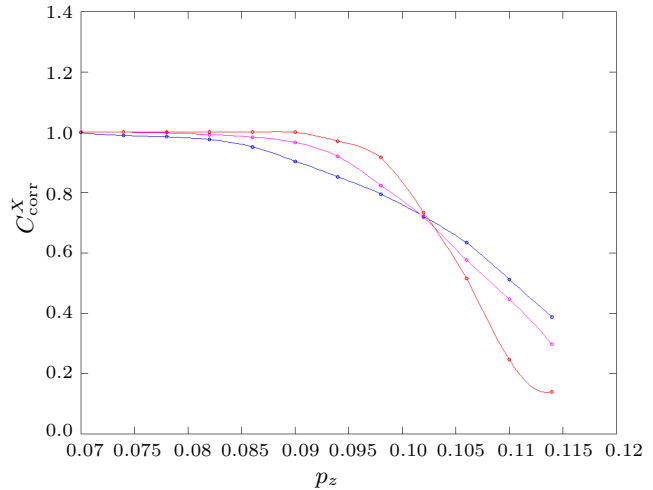

Employing this algorithm, we obtain the decay of the autocorrelation function $C^Z_{\m{corr}}$ as a function of $p_x$ illustrated for different lattice sizes in \figref{fig:decay}.
The curves for different lattice sizes intersect for $p_c \gtrsim 0.102$, so that in the limit of $L\ra\infty$ error correction is possible for $p_x , p_y < p_c$.
Our numerically obtained value is only slightly smaller than the theoretical value $p_c \simeq 0.1094 \pm 0.0002$ \cite{Dennis02}.
For the uncorrected or `bare' autocorrelation functions we have $2^{-N}\tr\left[X\cdot\Phi_{\m{err}}\mdag(X)\right]\approx0$ in the whole parameter regime depicted in \figref{fig:decay}, 
so there is a regime where our error-correction procedure is maximally beneficial, bringing the autocorrelation function from $0$ to $1$.

\subsection{Error model}

We now turn to the situation we are actually physically interested in, namely where $\Phi_{\m{err}}^t$ is induced by the memory Hamiltonian $H$ and coupling of the memory to a thermal environment.
We consider a Davies weak-coupling limit \cite{Davies74} and briefly summarize its discussion in \cite{Chesi10,Haah11c,Beat12}. In this limit, the evolution of the memory is described by a Markovian Master equation
\begin{align}
\dot{\rho_t} = -i\left[H , \rho_t\right] + \cL(\rho_t)
\end{align}
where the interaction between the memory and the bath is captured in the unitarity-breaking Lindblad operator $\cL$.
We assume that the environment is weakly coupled to the bath through single-qubit Pauli operators and thus is able to induce spin- and phase-flip errors, leading to transitions between eigenstates of $H$ that differ only by the application of a single-qubit Pauli operator.
Processes in which an energy $\omega$ is transferred from the anyonic system to the bath happen with rate $\gamma(\omega)$, which depends on how the bath is modeled.
An expression for $\gamma(\omega)$ often found in the literature is given by
\begin{equation}\label{eq:gamma_bath}
\gamma(\omega)=2 \kappa_n \left|\frac{\omega^n}{1-e^{-\beta\omega}}\right|
e^{-|\omega|/\omega_c}
\end{equation}
and can be derived from a spin-boson model \cite{Leggett87, DiVincenzo05}.
In the following, we set the cutoff frequency of the bath $\omega_c \to \infty$ for simplicity. A bath with $n = 1$ is called `Ohmic', whereas one with
$n \geq 2$ is called `super-Ohmic'. Only the former case is considered in the numerical simulations \cite{super_comment}.
To summarize, we will use
\begin{align}\label{eq:gamma_simple}
 \gamma(\omega)=\gamma(0) \cdot \left|\frac{\beta\omega}{1-e^{-\beta\omega}}\right|
\end{align}
where we think of $1/\gamma(0) = (2\kappa_1T)^{-1}$ as the relevant microscopic timescale, since the diffusion of anyons is widely determined by $\gamma(0)$.
Note that \eqref{eq:gamma_bath} and \eqref{eq:gamma_simple} fulfill the detailed balance condition $\gamma(-\omega) = \gamma(\omega)\cdot e^{-\beta\omega}$, guaranteeing that the Gibbs state is the fixed point of the Markovian dynamics,
$\cL\left(e^{-\beta H}\right) = 0$.

In such a physical model, the weight of a hypothetical error chain is not simply given by its Manhattan length, as it was the case in \eqref{eq:indep_weight}. 
It is in general not true that the most likely error chain is the one with the smallest number of spin flips. 
The number of hoppings some time $\Delta t$ after the creation of an anyon or a pair of anyons is Poisson-distributed and for large $\gamma(0)\cdot\Delta t$ there is in fact a \emph{small} probability that the number of spin flips is still small. 
Using the Manhattan distance as the weight of an error chain connecting two anyons (as done in \cite{Chesi10,Beat12}) seems thus hard to justify.
Rather than trying to find the most likely error chain, we therefore try to find the most likely \emph{pairing} of the defects (a pairing may either be between two anyons or between an anyon and a boundary). 
This is not exactly equivalent to finding the most likely equivalence class of errors but should not make a relevant difference in practice and is numerically feasible. 

\begin{figure}
	\includegraphics[width=0.5\textwidth]{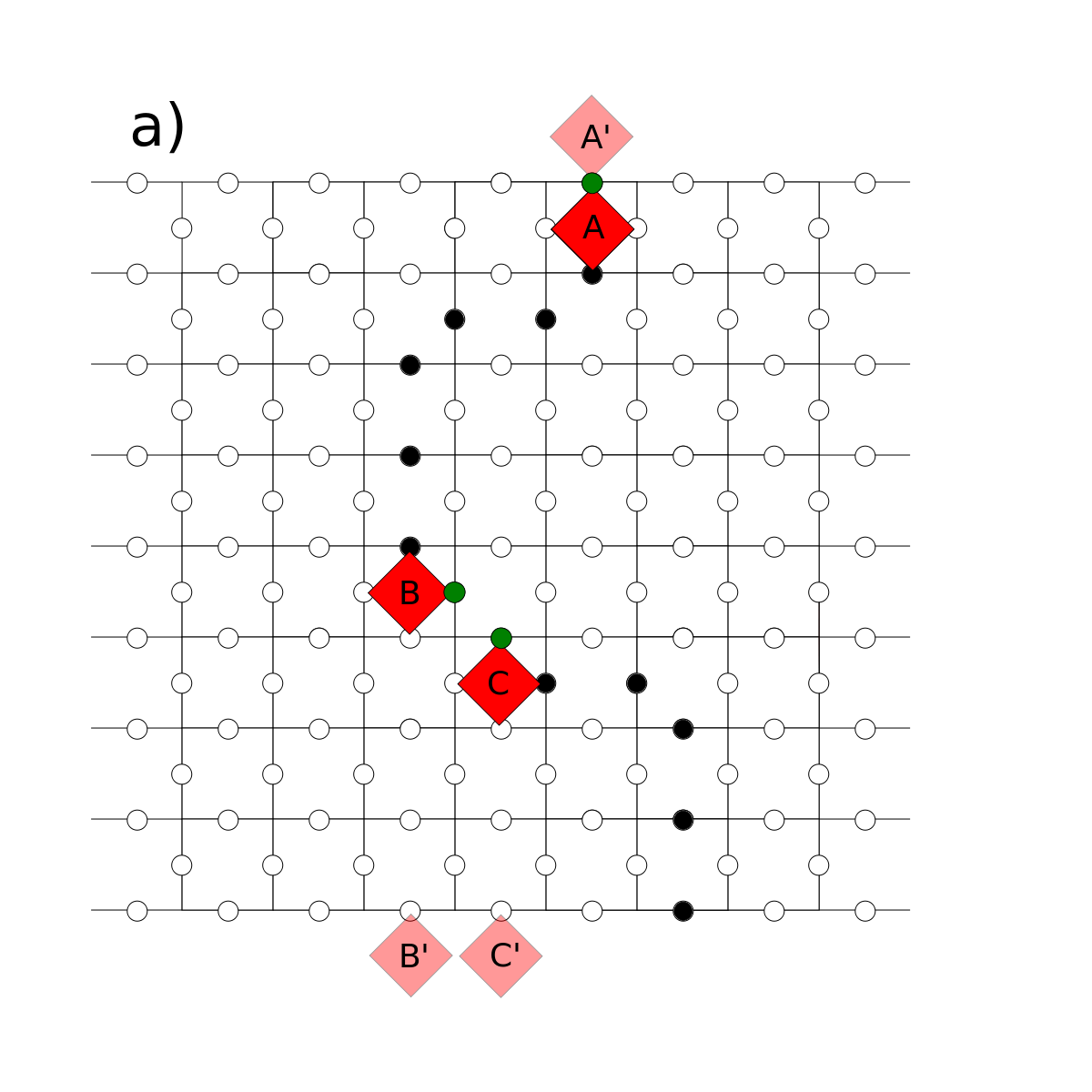}
	\includegraphics[width=0.23\textwidth]{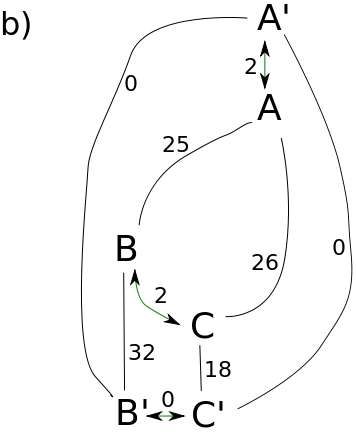}
	\caption{The code has suffered $\sigma_x$ errors at the black qubits in a). The syndrome measurement detects plaquette anyons $A$, $B$, and $C$. 
	For the error correction procedure, we first perform a Delaunay triangulation of the full graph with edges connecting all real anyons (in the case of only three real anyons, the full graph and the triangulation coincide).
	Then, virtual anyons $A'$, $B'$, and $C'$ are added.
	The graph b) is then used for the perfect matching algorithm, where the obtained matching is highlighted with arrows. Error correction fails, because anyons $B$ and $C$ are fused and anyon $A$ is moved to the upper boundary,
        thereby performing a logical $X$ operator on the qubit stored in $\kil_0$.}
	\label{fig:matching}
\end{figure}

The random walk of an anyon on the grid leads to a diffusive spreading of the probability distribution. The probability of finding it at time $t$ at a position $\vec{r}$ relative to its position at time $t'$ is 
\begin{align}\label{eq:diffusion}
 \frac{1}{4\pi D(t-t')} \cdot e^{-\vec{r}^2/4D(t-t')}\ .
\end{align}
In the case of an Ohmic bath, the diffusion constant $D$ is basically given by the hopping rate $\gamma(0)$ \cite{Chesi10}.
Similarly, the distance vector of two anyons that have been jointly created diffuses with a constant $2D$ since both of its ends are moving.
A sensible choice for the weight of an edge between two anyons is therefore the square of their Euclidean distance, while for an edge connecting an anyon to its closer boundary we take the weight to be \emph{twice} the square of the Euclidean distance.
A more thorough justification of this choice can be found in Appendix \ref{app:diffusion}. 
An example for the application of the error correction algorithm is given in \figref{fig:matching}.

\section{Long-range repulsion between anyons}\label{sec:longrange}

A $1/r^\alpha$ repulsive potential with $0\leq\alpha<2$ between the anyons allows one to increase the lifetime of the toric code arbitrarily by increasing $L$ \cite{Chesi10}.
We study the $\alpha=0$ case here, since for this case proposals of its physical implementation exist \cite{Chesi10,Pedrocchi11}.
Furthermore, this case can be analyzed analytically without need for a mean-field approximation and provides the best scaling of the lifetime with $L$.
Since the repulsive potential is independent of the anyon distances $r$, its numerical simulation has the lowest cost.

We study a system with a repulsion between the anyons which is spatially constant, so the total Hamiltonian is
\begin{align}\label{eq:constInter}
 H = H_{\m{Kitaev}} + \frac{A}{2}\cdot\left(\sum_{p\neq p'}n_pn_{p'}+\sum_{s\neq s'}n_sn_{s'}\right)\ ,
\end{align}
where $H_{\m{Kitaev}}$ is as in \eqref{eq:Kitaev} \cite{3qubit_comment}.
Basically, the total energy is now parabolically rather than linearly increasing in the anyon numbers.
The physical realization of such an interaction and its effect on the lifetime of the toric code have been studied in detail in \cite{Chesi10,Pedrocchi11,Beat12}.

Let us first discuss the toric code, i.e.\ a memory with periodic boundary conditions. The lifetime of the memory is given by the time when finding the most likely anyon pairing becomes ambiguous.
After a time $t$, the distance between the two anyons of a pair is of order $\sqrt{Dt}$, with a diffusion constant $D$.
Let $n_{\m{eq}}$ be the equilibrium density of anyons, such that their average separation is $\sim 1/\sqrt{n_{\m{eq}}}$. 
We may then estimate the lifetime of the memory as the time when the anyons have diffused over their average distance and error correction becomes ambiguous,
\begin{align}\label{eq:lifetime}
 \tau(\eps) = \frac{c(\eps)}{D\cdot n_{\m{eq}}}\ .
\end{align}
We use $c(\eps)$ as a single fitting parameter, which can be thought of as a critical fraction of spins affected by errors and will be of order of a few percents \cite{Chesi10}.

Let 
\begin{align}\label{eq:eeq}
 e_{\m{eq}} = \Delta + A (L^2 n_{\m{eq}} - 1)
\end{align}
denote energy per anyon in equilibrium. 
Since there is either one or no anyon at each position, the equilibrium anyon density can be determined self-consistently from
\begin{align}\label{eq:neq}
 n_{\m{eq}} = \left[\exp(\beta e_{\m{eq}}) + 1\right]^{-1}\ .
\end{align}
The diffusion constant is $D = \gamma(0) + 4\gamma(-2e_{\m{eq}})$, which for an Ohmic bath is widely dominated by the first summand \cite{Chesi10}.

Straightforward algebra (c.f.\ Appendix \ref{app:estimates}) gives two simple bounds on $n_{\m{eq}}$ for large enough $L$. We have that
\begin{align}\label{eq:bound1}
 n_{\m{eq}} > \frac{1}{L^2} \quad \m{if} \quad L^2 > \exp(\beta\Delta) + 1
\end{align}
and
\begin{align}\label{eq:bound2}
 n_{\m{eq}} < \frac{1}{L^{2-\eps}} \quad \m{if} \quad \frac{\ln L}{L^\eps-1} < \frac{\beta A}{2-\eps}\ .
\end{align}
Putting these bounds into \eqref{eq:neq} we also find
\begin{align}\label{eq:eeqBound}
 L^{2-\eps} < \exp(\beta e_{\m{eq}}) + 1 < L^2
\end{align}
for large enough $L$.
The lifetime \eqref{eq:lifetime} is inverse in $n_{\m{eq}}$ and will thus for any $\eps>0$ grow faster than $L^{2-\eps}$ as $L \ra \infty$. 

\begin{figure}
  \setlength{\unitlength}{0.5\textwidth}
    \begin{picture}(0.8,0.8)	
	\put(0.02,0.09){\includegraphics[width=0.4\textwidth]{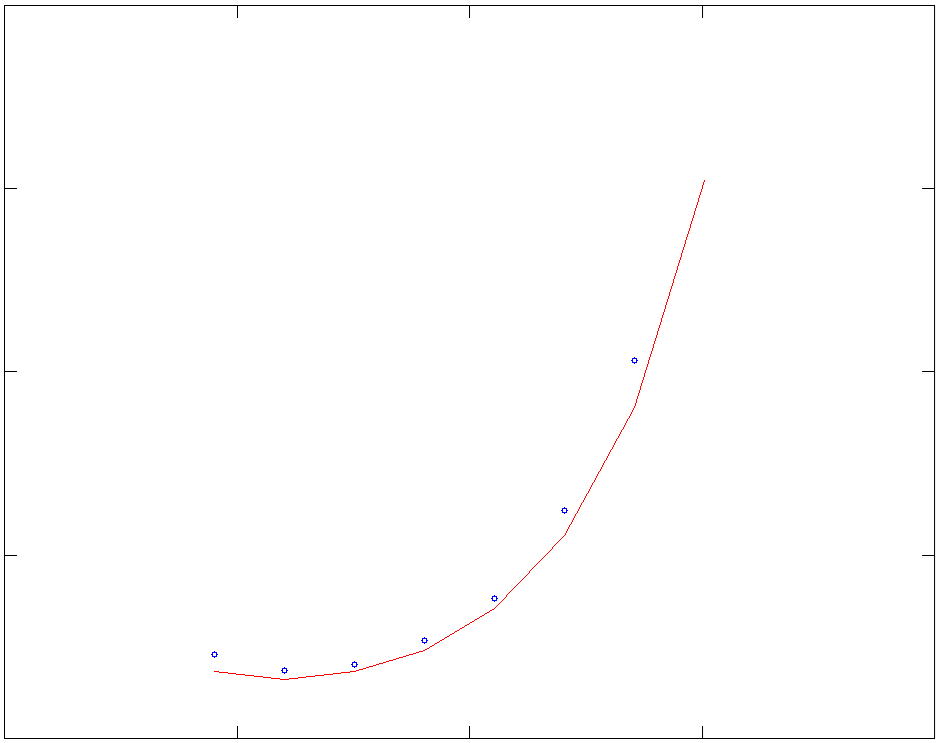}}

	\put(-0.06,0.36){\begin{sideways}ratio\end{sideways}}
	\put(0.36,0.02){size}

	\put(0.0,0.06){\footnotesize $10^0$}
	\put(0.2,0.06){\footnotesize $10^1$}
	\put(0.4,0.06){\footnotesize $10^2$}
	\put(0.6,0.06){\footnotesize $10^3$}
	\put(0.8,0.06){\footnotesize $10^4$}

	\put(-0.01,0.090){\footnotesize $0$}
	\put(-0.01,0.2450){\footnotesize $5$}
	\put(-0.02,0.400){\footnotesize $10$}
	\put(-0.02,0.555){\footnotesize $15$}
	\put(-0.02,0.710){\footnotesize $20$}

    \end{picture}
  \caption{Ratio of the number of anyons created on the boundary to the number of anyons created in the bulk if the anyon population approaches its equilibrium value. The blue circles show numerical results sampled over a time $10^5 \cdot (\kappa_1\Delta)^{-1}$. 
		 The red curve shows the analytical prediction $\exp(\beta e_{\m{eq}}) / 4L$. We have used parameters $\beta\Delta = 1/0.3$ and $A/\Delta = 0.1$ and a bath $\gamma(\omega)$ as in \eqref{eq:gamma_simple}.}
  \label{fig:ratio}
\end{figure}
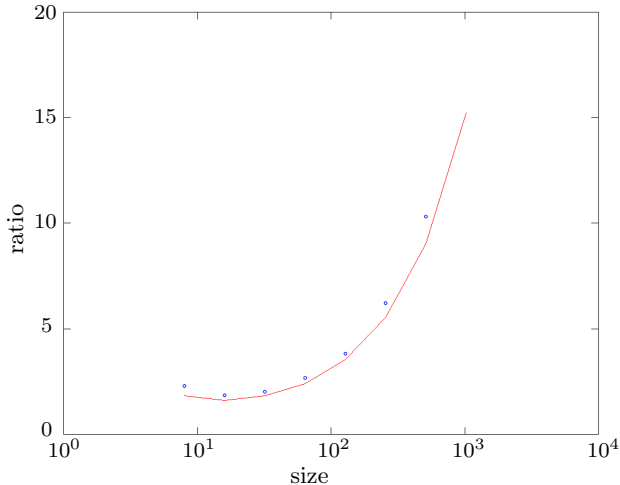

In the case of the planar code, single anyons can be created at and absorbed by the boundaries. The anyon production rate per boundary spin is $\gamma(-e_{\m{eq}})$ (if the anyon density approaches the equilibrium density), while it is $2\gamma(-2e_{\m{eq}}-A)$ for spins in the bulk. 
Again, $e_{\m{eq}}$ can be determined self-consistently from \eqref{eq:eeq} and \eqref{eq:neq}, where we replace $L^2$ by $L(L+1)$ in \eqref{eq:eeq} for the planar code.
In order to find the total creation rates, these single-spin production rates have to be multiplied by $2(L+1)$ and $2L^2-1$, the number of spins on the boundary and in the bulk, respectively.
The total rate for production of anyons on the boundaries is then, using the bath \eqref{eq:gamma_simple},
\begin{align}\label{eq:bdryRate}
 2(L+1)\cdot\gamma(-e_{\m{eq}}) \simeq 2(L+1)\frac{\beta e_{\m{eq}}}{\exp(\beta e_{\m{eq}})}\gamma(0)\ ,
\end{align}
while the total rate for the production of anyons in the bulk is
\begin{align}\label{eq:bulkRate}
 (2L^2-1)\cdot2\gamma(-2e_{\m{eq}}-A) \simeq (2L^2-1)\frac{4\beta e_{\m{eq}}}{\exp(2\beta e_{\m{eq}})}\gamma(0)\ .
\end{align}
Since the anyon density vanishes for large enough $L$, we neglected for these estimates that in fact only those spins that have no adjacent anyons should be considered for the anyon production rate.
Using \eqref{eq:eeqBound} with $\eps=\half$, we see that both the anyon production rate on the boundaries and in the bulk go to $0$ as $L\ra\infty$.
But which of the two will dominate for large $L$?
Applying again \eqref{eq:eeqBound}, it is clear that the ratio $\simeq \frac{\exp(\beta e_{\m{eq}})}{4L}$ of the boundary to the bulk rate is in fact \emph{diverging} with $L$. 
The analytics are in excellent agreement with numerical simulations (\figref{fig:ratio}) that show an increasing ratio after a slight minimum at $L \simeq 16$.

Let us note that the ratio of ``active'' anyons is in fact slightly different than what the above analysis suggests.
The probability that an anyon created at the boundary is not reabsorbed before it gets to the second row is $\frac{\gamma(0)}{\gamma(0)+\gamma(+e_{\m{eq}})} \simeq \frac{\gamma(0)}{\gamma(+e_{\m{eq}})}$.
Similarly, the probability that a newly created pair is not immediately re-annihilated is $\simeq \frac{6\gamma(0)}{\gamma(2e_{\m{eq}}+A)}$.
If we are only interested in the ratio of boundary to bulk anyons that ever get away from their place of creation, we thus obtain a further factor $\frac{\gamma(2e_{\m{eq}}+A)}{6\gamma(+e_{\m{eq}})}$.
For the bath \eqref{eq:gamma_simple}, this factor converges to $\frac{1}{3}$ for large enough $L$. The ratio of ``active'' anyons created at the boundaries to ``active'' anyons created in the bulk is thus finally $\frac{\exp(\beta e_{\m{eq}})}{12L}$.

An anyon created at the boundary has three possible fates. 
\begin{enumerate}
 \item It can be reabsorbed by the same boundary (which will happen to most of them) or fuse with an anyon created at the same boundary, leading to a trivial operation performed on the degenerate ground space.
 \item It can fuse with an anyon created at the opposite boundary or be absorbed by the opposite boundary, leading to a logical error.
 \item It can fuse with an anyon created in the bulk.
\end{enumerate}
We will argue that the time it takes to create a logical error via possibility 2 is larger than the lifetime of the memory in the toric case and that possibility 3 has a beneficial effect on the lifetime of the memory.
Combining these arguments with the fact that anyon production on the boundary surpasses the anyon production in the bulk, we conclude that the boundary has a positive net effect on the lifetime of the memory.

Let us study the time it takes to create a logical error through possibility 2.
There are two time-scales involved in the creation of such an error: The time it takes to create an anyon that walks during its lifetime at least once to the opposite half of the grid and the time it takes to do this walk.
What is the probability that an anyon created at a boundary ever moves to the opposite half of the grid and what is the expected number of hoppings necessary for this?
Up to the boundary conditions, this problem is exactly equivalent to a well-known mathematical problem called The Gambler's Ruin. 
If a newly created anyon is to ever reach the opposite half, it first has to avoid immediate reabsorbtion. The probability of ever getting to the second row is $\simeq \frac{\gamma(0)}{\gamma(+e_{\m{eq}})}$. 
Now imagine that the distance of the anyon to the boundary that has created it represents a gambler's bankroll. He starts with one unit of money (the anyon starts at the first row) and then does a series of fair coin flips for one unit of money each.
He ends when either he gets broke (the anyon gets reabsorbed by the same boundary that has created it) or his bankroll reaches $L/2$ units of money (the anyon reaches the opposite half of the grid).
Since the gambler starts with one unit of money and he does a series of games with zero expectation value, his probability of ever reaching $L/2$ has to be $2/L$, which also answers the corresponding question for the anyon.
Note that movements of the anyon parallel to the boundary that has created it are irrelevant for the production of logical errors, which is why the problem can be mapped to such a one-dimensional one. 
Note that in the case of anyons this probability provides in fact an upper bound, since we neglected the possibility that the anyon fuses with another one before reaching the opposite half of the grid.
A formally correct treatment of the problem can be found in Appendix \ref{app:gambler}. There, we also show that the average number of coin flips (hoppings perpendicular to the creating boundary) needed to reach $L/2$ is $(L/2)^2$.

To summarize, anyons are created at the boundaries with rate $2(L+1)\gamma(-e_{\m{eq}})$, only a fraction $\frac{\gamma(0)}{\gamma(+e_{\m{eq}})}$ do not immediately get reabsorbed and only a fraction of at most $\frac{2}{L}$ ever reaches the opposite half of the grid.
If an anyon does reach the opposite grid, this takes on average a time $\frac{1}{\gamma(0)}\cdot2\cdot\left(\frac{L}{2}\right)^2$ (the factor $2$ takes hoppings parallel to the creating boundary into account). The total lifetime of a memory in which anyons are only created at the boundaries may therefore be estimated as
\begin{align}\label{eq:bdryLifetime}
 &\frac{1}{2(L+1)\gamma(-e_{\m{eq}})}\cdot\frac{\gamma(+e_{\m{eq}})}{\gamma(0)}\cdot\frac{L}{2} + \frac{1}{\gamma(0)}\cdot\frac{L^2}{2} \nn\\
 &\quad\simeq \frac{1}{4\gamma(0)}\cdot\left(e^{\beta e_{\m{eq}}}+2L^2\right)
\end{align}
where we used detailed balance. Note that the only time-scale that entered the first summand was the anyon creation time $1/\gamma(-e_{\m{eq}})$. 
However, this cancels exactly with $\gamma(+e_{\m{eq}})$: the higher the anyon production rate, the higher is (by detailed balance) the probability that an anyon is immediately reabsorbed, such that the only remaining time-scale is $1/\gamma(0)$.

The only property of the bath $\gamma(\omega)$ we used was the detailed balance property, so this scaling behavior is independent of the particulars of the bath and not specific for \eqref{eq:gamma_simple}.
However, in the case of a super-Ohmic bath with $\gamma(0)=0$ the analysis of the Gambler's Ruin problem has to be redone with an effective hopping rate emerging from indirect hopping processes that scale with $\gamma(-2e_{\m{eq}}-A)$ \cite{Chesi10}.
Since $e_{\m{eq}}$ diverges logarithmically with $L$ (\ref{eq:eeqBound}), the effective hopping rate becomes vanishing for large $L$, leading to an improved scaling of the memory lifetime with $L$ \cite{Chesi10}. 

Using \eqref{eq:neq} we see that the first summand in \eqref{eq:bdryLifetime}, i.e.\ the timescale needed for the creation of an anyon that walks to the opposite half, is for large enough $L$ identical to the lifetime of the toric code \eqref{eq:lifetime}.
Furthermore, we know from $\eqref{eq:eeqBound}$ that the second summand grows faster than the first one. 
We conclude that the time needed for anyons created on the boundaries to produce a logical error grows faster with $L$ than the time before the matching of anyons produced in the bulk becomes ambiguous.

In the limit of very large $L$, anyons are created almost exclusively on the boundary, once the anyon production rates have approached their equilibrium values. 
However, as long as the anyon population is small enough anyon production in the bulk will outweigh anyon production on the boundaries, so that possibility 3 is non-negligible even for large $L$.

Possibility 3 has the same effect as if the bulk anyon with which the boundary anyon fused had moved to and been absorbed by the creating boundary.
The effect of possibility 3 is therefore to cause an effective bias of bulk anyons to move to the closer boundary.
We saw that the probability that a boundary anyon gets a certain distance away from its creating boundary decreases at least inversely with that distance.
This effect is thus the stronger, the closer the bulk anyons are to the boundary. 
We saw in the discussion of the toric case that error correction breaks down if the anyons that have been created as parts of the same pair have moved sufficiently far away from each other such that the pair-matching becomes ambiguous.
If two such anyons, however distant they are, now move to the same boundary this ambiguity is resolved without a logical error. 

In conclusion, the equilibrium density of anyons $n_{\m{eq}}$ is not affected by the boundary conditions, but the fraction of anyons created on the boundary becomes dominant for large enough $L$. 
The time it takes for boundary anyons to create a logical error grows faster with $L$ than the lifetime of the toric code. 
We expect a beneficial effect from the possibility of boundary anyons fusing with bulk anyons. 
We conclude that for large enough $L$ a planar code of size $L$ has a larger lifetime than a toric code of size $L$.

\begin{figure}
  \setlength{\unitlength}{0.5\textwidth}
    \begin{picture}(0.8,0.8)	
	\put(0.05,0.10){\includegraphics[width=0.4\textwidth]{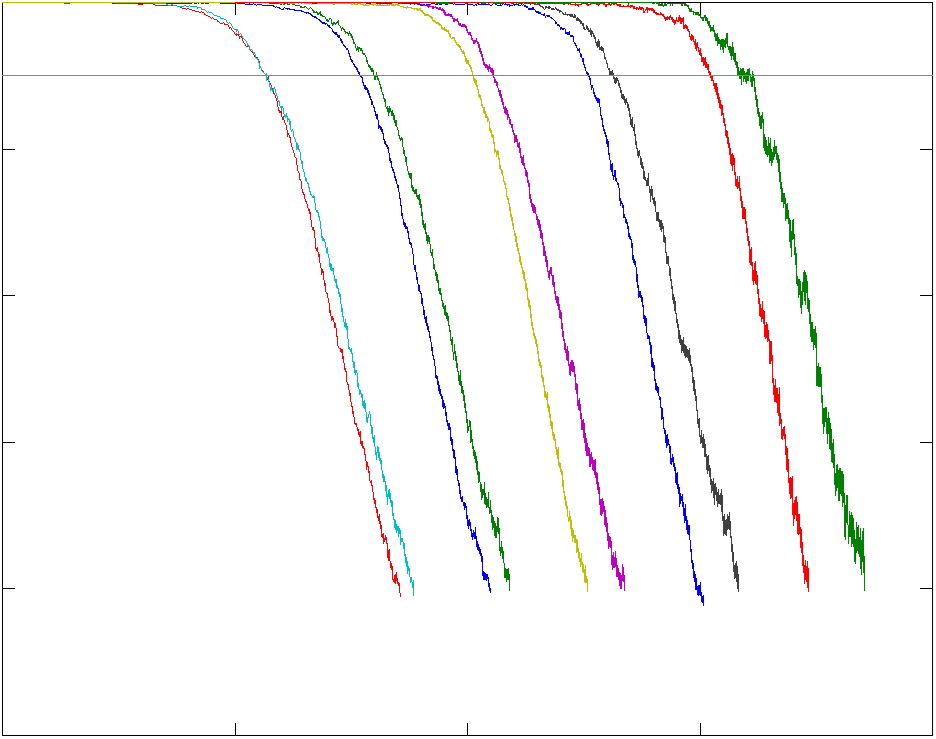}}
	
	\put(-0.06,0.36){\begin{sideways}$C^Z_{\m{corr}}(t)$\end{sideways}}
	\put(0.28,0.02){$t \,\,\, \left(\text{units of} \,\,\, (\kappa_1\Delta)^{-1}\right)$}

	\put(0.03,0.07){\footnotesize $10^0$}
	\put(0.23,0.07){\footnotesize $10^1$}
	\put(0.43,0.07){\footnotesize $10^2$}
	\put(0.63,0.07){\footnotesize $10^3$}
	\put(0.83,0.07){\footnotesize $10^4$}

	\put(-0.005,0.10){\footnotesize $0.0$}
	\put(-0.005,0.225){\footnotesize $0.2$}
	\put(-0.005,0.350){\footnotesize $0.4$}
	\put(-0.005,0.475){\footnotesize $0.6$}
	\put(-0.005,0.600){\footnotesize $0.8$}
	\put(-0.005,0.725){\footnotesize $1.0$}

    \end{picture}
  \caption{Temporal decay of the autocorrelation functions $C^Z_{\m{corr}}(t)$ for different lattice sizes and boundary conditions, for the Hamiltonian \eqref{eq:constInter} and the bath \eqref{eq:gamma_simple}. 
	    Times are in units of $(\kappa_1\Delta)^{-1}$ and the physical parameters are $\beta\Delta = 1/0.3$ and $A/\Delta = 0.1$. From left to right (at height $0.6$, say) we have (grid type -- size $L$) toric 32, planar 32, toric 64, planar 64, toric 128, planar 128,
	    toric 256, planar 256, toric 512, and planar 512. The intersection of $C^Z_{\m{corr}}(t)$ with the horizontal line at height $0.9$ is used to determine the lifetimes $\tau(0.1)$ given in \figref{fig:lifetimes}.}
  \label{fig:autocorr}
\end{figure}

\begin{figure}
  \setlength{\unitlength}{0.5\textwidth}
    \begin{picture}(0.8,0.8)	
	\put(0.05,0.1){\includegraphics[width=0.4\textwidth]{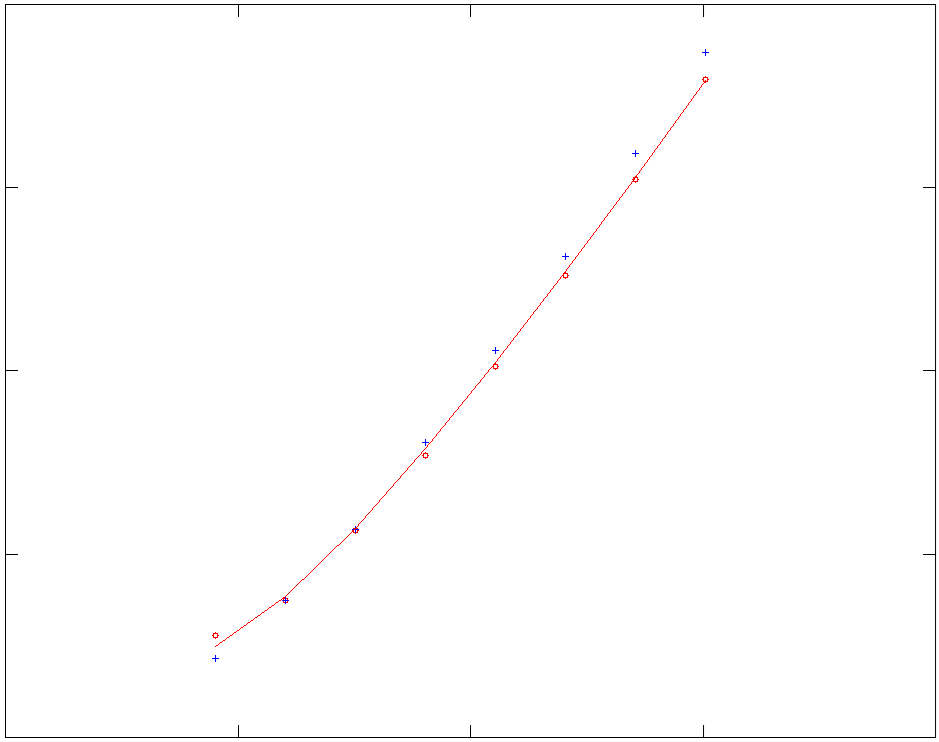}}
	
	\put(-0.06,0.2){\begin{sideways}$\tau(0.1) \,\,\, \left(\text{units of} \,\,\, (\kappa_1\Delta)^{-1}\right)$\end{sideways}}
	\put(0.36,0.02){size ($L$)}

	\put(0.03,0.07){\footnotesize $10^0$}
	\put(0.23,0.07){\footnotesize $10^1$}
	\put(0.43,0.07){\footnotesize $10^2$}
	\put(0.63,0.07){\footnotesize $10^3$}
	\put(0.83,0.07){\footnotesize $10^4$}

	\put(-0.005,0.10){\footnotesize $10^0$}
	\put(-0.005,0.255){\footnotesize $10^1$}
	\put(-0.005,0.41){\footnotesize $10^2$}
	\put(-0.005,0.565){\footnotesize $10^3$}
	\put(-0.005,0.72){\footnotesize $10^4$}

    \end{picture}
  \caption{Lifetimes $\tau(0.1)$ of the quantum information stored in the systems described in the caption of \figref{fig:autocorr}. The blue crosses represent planar codes, the red circles toric codes. 
		  The sizes $L$ of the codes are 8, 16, 32, 64, 128, 256, 512, and 1024.
		The red line shows the analytical prediction for the lifetime in the toric case obtained from \eqref{eq:lifetime} with $c(0.1) = 5.1\%$.
		\cite{Chesi10} find with an analogous plot [Fig.\ 6] $c(0.1)=4.4\%$. The increased lifetime is due the choice of the square of the Euclidean distance rather than the Manhattan length as the weight of an error chain.}
  \label{fig:lifetimes}
\end{figure}

This is indeed confirmed by our numerical simulations. Already for $L \gtrsim 32$ the lifetime $\tau(0.1)$ of the planar code exceeds the one of the toric code. 
For very small memories ($L < 10$) the toric code is superior, which may be attributed to logical errors caused by single anyons. 
In the planar code in \figref{fig:stabilizers}, for example, an anyon created at a boundary needs to perform only $4$ hoppings to cause a logical error when error correction is performed. 
\figref{fig:autocorr} illustrates the temporal decay of the stored quantum information by depicting the autocorrelation functions $C^Z_{\m{corr}}(t)$ for different lattice sizes and for both planar and toric grids. 
The obtained lifetimes are illustrated in \figref{fig:lifetimes}.

While we included the analytical prediction \eqref{eq:lifetime} for the lifetime of the toric code in \figref{fig:lifetimes}, we cannot give such a simple expression for the lifetime of the planar code.
When $L$ is increased, the effect of the boundary changes from adversarial to beneficial.

\section{The honeycomb model as a quantum memory}\label{sec:honeycomb}

\subsection{From the honeycomb to the planar code}

The toric code Hamiltonian \eqref{eq:Kitaev} (involving local \emph{four}-qubit couplings) can be realized as a low-energy effective Hamiltonian of the Kitaev honeycomb model, which involves only local \emph{two}-qubit couplings \cite{Kitaev06,Vidal08,Pedrocchi11}.
The Hamiltonian of the honeycomb model can be simplified through a spin to hard-core boson transformation \cite{Vidal08}. 
In this new language, the toric code Hamiltonian \eqref{eq:Kitaev} emerges as a low-energy (no hard-core bosons present) fourth-order effective Hamiltonian. 
The qubits of the toric code obtained this way are not identical to the physical qubits of the underlying honeycomb lattice, but are \emph{effective} qubits.
The interactions of \eqref{eq:Kitaev} between four of these effective qubits are mediated through processes in which two pairs of virtual hard-core bosons are created and then fuse again to the vacuum.
An interaction involving three effective qubits would correspond to a process in which a pair of virtual hard-core bosons is created from the vacuum, followed by a hopping and an annihilation process.
Clearly, such processes give no contribution to the effective Hamiltonian and the third-order effective Hamiltonian vanishes.
There is no way of obtaining from the honeycomb model the three-qubit boundary terms 
introduced in \cite{Bravyi98} and used so far in this paper. 

\begin{figure}
	\includegraphics[width=0.5\textwidth]{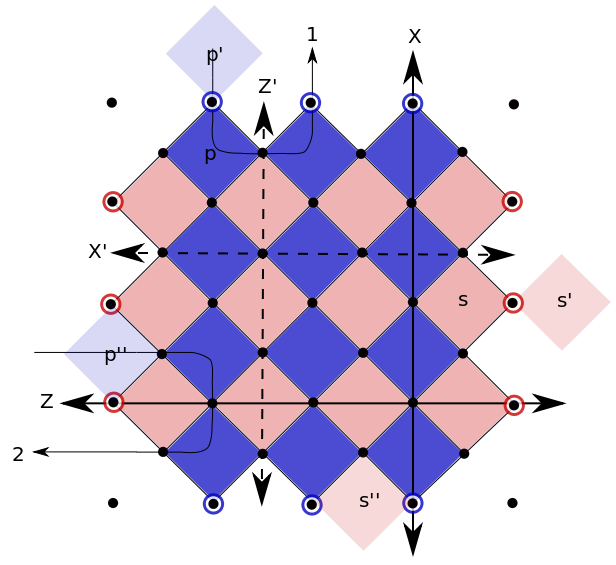}
	\caption{An effective planar code obtained from a honeycomb lattice with a boundary. The black dots represent effective qubits emerging from the underlying honeycomb model.
	The four isolated effective qubits in the corners are depicted for completeness but not needed for information storage. 
	Blue (dark) squares (like $p$) are plaquette operators, red (light) squares (like $s$) are star operators.
	There are two non-equivalent ways of defining a pair of macroscopic Pauli-like observables, the ``big'' logical operators $X$ and $Z$ and the ``small'' logical operators $X'$ and $Z'$. 
	However, an undetectable logical error consisting of only three single-qubit errors can be performed on all of these operators. 
	For instance, a sequence of $\sigma_x$ errors on the three qubits along path $1$ ($2$) causes an error on the logical $Z'$ ($Z$) operator.}
	\label{fig:checkboard}
\end{figure}

\figref{fig:checkboard} schematically shows a planar code obtained from the honeycomb model, involving only four- but no three-qubit operators.
The ground space of the Hamiltonian obtained this way, i.e.\ the space stabilized by all four-qubit operators $(\sigma_z)^{\otimes4}$, $(\sigma_x)^{\otimes4}$, 
has then a large degeneracy growing with the size of the memory, even if we forget about the four isolated qubits in the corners that do not interact with anything.

Let $\group$ denote the group generated by all four-qubit stabilizers.
In an abstract language, elements of the Pauli group on all effective qubits (except the four isolated ones in the corners) that are in the centralizer $C(\group)$ of the stabilizer group $\group$, 
but not in $\group$ itself, perform logical operations on the stabilized subspace that cannot be detected through violation of a stabilizer. 
The goal is then to find Pauli-like observables $X, Z \in C(\group)\setminus\group$ which allow us to store a qubit that is topologically protected. 
Elements of  $C(\group)\setminus\group$ that commute with $X$ and $Z$ will not do any harm to the stored qubit, but those that do not commute with $X$ or $Z$ can. For topological protection we therefore want
\begin{itemize}
 \item $X$ and $Z$ to have a macroscopic distance (i.e.\ to necessitate $O(L)$ single-qubit operations), and
 \item all elements of $C(\group)\setminus C(\left\langle X,Z \right\rangle)$ to have macroscopic distance.
\end{itemize}
Unfortunately, in the planar code obtained from the honeycomb model there is no choice of logical operators $X$ and $Z$ that fulfill these requirements.
As illustrated in \figref{fig:checkboard}, the pairs $X, Z$ and $X', Z'$ fulfill the first requirement but not the second one.

We will therefore expand $\group$ in such a way that all elements of $C(\group)\setminus\group$ have a macroscopic distance.
For the space stabilized by $\group$ being non-trivial we need $\group$ to be Abelian and to not contain $-I$.
All of these requirements are met if we add to $\group$ the single-qubit operators $\sigma_x$ acting on all qubits surrounded by a red circle in \figref{fig:checkboard} and the operators $\sigma_z$ acting on all qubits surrounded by a blue circle.
Error paths $1$ and $2$ in \figref{fig:checkboard} now no longer are elements of $C(\group)\setminus\group$: 
$1$ is no longer an element of $C(\group)$ since it anti-commutes with two blue single-qubit operators in $\group$ while $2$ becomes an element of $\group$ (it is the product of the two red operators enclosed by it).
Therefore, the ``small'' operators $X'$ and $Z'$ now satisfy our requirements for topological protection while the ``big'' operators $X$ and $Z$ still violate the second requirement.

Our new stabilized subspace topologically protects exactly one qubit. 
In fact, our new code is exactly equivalent to the planar code with three-qubit operators on the boundary, since the single-qubit stabilizer operators effectively eliminate the two degrees of freedom of the qubits on the boundary.
However, the Hamiltonian dynamics are different since there is no energy penalty associated with the violation of the single-qubit stabilizers on the boundary. 
We therefore assume that we are able to perform the measurements corresponding to the single-qubit stabilizers at the read-out step. 
Note that these operators act on \emph{effective} qubits, but $\sigma_x$ and $\sigma_z$ measurements performed on them indeed correspond to measurements of a single \emph{physical} qubit of the underlying honeycomb lattice. We refer to \cite{Pedrocchi11} for details about the mapping between physical and effective qubits.

If one of the single-qubit stabilizers on the boundary is violated, we may imagine that an anyon is present at the position adjacent to it which is outside of the actual grid.
For example, a $\sigma_x$ ($\sigma_z$) error can create two plaquette (star) anyons at positions $p$ and $p'$ ($s$ and $s'$) in \figref{fig:checkboard}. An anyon at position $p'$ ($s'$) can only escape to $p$ ($s$).
In terms of anyon dynamics there are now for both kinds of anyons two boundaries that can create and absorb them and two boundaries that ``attract'' and store them.
For example, a plaquette (star) anyon at position $p$ ($s$) can reduce its energy by hopping to position $p'$ ($s'$). Analogously, escaping from one of these boundary positions has an energy cost.
An anyon that hops to position $p''$ or $s''$ has been absorbed by the corresponding boundary and can, unlike one at position $p'$ or $s'$, no longer be detected through a violated stabilizer operator.
During error correction, an anyon stored at position $p'$ or $s'$ will be moved to the interior of the grid.

\subsection{Repulsion between anyons and anyon holes}

As discussed, the simple toric code Hamiltonian \eqref{eq:Kitaev} does not provide a lifetime of the stored quantum information that can be increased by making the memory larger. 
Pedrocchi \emph{et al}.\ \cite{Pedrocchi11} studied a honeycomb model as introduced above coupled to cavity modes.
The cavity modes allow the read-out of the error syndrome of the effective toric code through frequency shifts. 
In a resonant parameter regime, the Hamiltonian 
\begin{align}\label{eq:holeRepulsion}
  H_{\m{eff}} = \Delta \sum_{a, a'} n_a \bar{n}_{a'}
\end{align}
is found perturbatively [eq.\ (52) in \cite{Pedrocchi11}].
Here, the sum $\sum_a$ runs over all stars and plaquettes (four-qubit operators only) and $\bar{n}_a = 1-n_a$ counts anyon holes. For $\Delta > 0$ this Hamiltonian describes an effective repulsion between anyons and anyon holes. 
For details about how $\Delta > 0$ can be achieved, we refer to \cite{Pedrocchi11} and Section VII in \cite{Chesi10}.

The requirement to stay strictly in the perturbative regime 
puts an upper bound $L^*$ on the linear size $L$ of the memory \cite{Pedrocchi11}.

With Hamiltonian \eqref{eq:holeRepulsion} and a total of $N$ four-qubit operators, the gap above the anyonic vacuum is $(N-1)\Delta$ and as long as $\sum_an_a < \frac{N}{2}$ the cost to add a further anyon is at least $\frac{N}{2}\Delta$.
If $L$ denotes the number of stars from top to bottom and the number of plaquettes from left to right (so $L=3$ in \figref{fig:checkboard}) we have $N=2L(L+1)$. 
With an anyon creation gap that grows quadratically in $L$ and a natural bath like \eqref{eq:gamma_simple} in which the creation rate decreases exponentially with the gap, 
we expect the production of anyon pairs in the bulk to be negligible against the creation of anyons at the boundaries. 
Note that if we say that a plaquette anyon is created at a boundary where single-qubit $\sigma_z$ measurements are performed, this implies that furthermore a virtual anyon is ``stored'' in this boundary.
From the discussion of the Gambler's Ruin problem in Appendix~\ref{app:gambler} it is clear that an anyon will do at most $\sim L^2$ hoppings before being absorbed by or stored in a boundary. 
We thus expect the time an anyon is present (and not stored in a boundary) to be much smaller than the time it takes to create a single anyon. To summarize, we are already for moderate values of $L$ in the regime
\begin{align}
 \frac{L^2}{\gamma(0)} \ll \left[4L\cdot\gamma\left((N-1)\Delta\right)\right]^{-1} \ll \left[L^2\cdot\gamma\left(2(N-2)\Delta\right)\right]^{-1}
\end{align}
where almost always one or no anyon is present. 

In this regime, error correction works as follows.
\begin{enumerate}
 \item Move any anyons in the interior of the grid perpendicularly to the closest of the four boundaries. (Each anyon is thereby moved to its most likely place of creation. We never assume that a pair of anyons has been created in the bulk.)
 \item If an anyon is ``stored'' in a boundary (at position $p'$ or $s'$ in \figref{fig:checkboard}, say), move it to the interior of the grid (to position $p$ or $s$).
 \item The anyons at each boundary can be matched with each other and the two-adjacent boundaries in two different not trivially suboptimal ways. Chose the one with lower weight, where the weight is given by the square of the Euclidean distance.
\end{enumerate}
For this system, the perfect matching problem could be reduced to a one-dimensional one. Both possible matchings that are not trivially suboptimal can be explicitly tested, such that no approximative algorithms are needed.

\begin{figure}
	\includegraphics[width=0.5\textwidth]{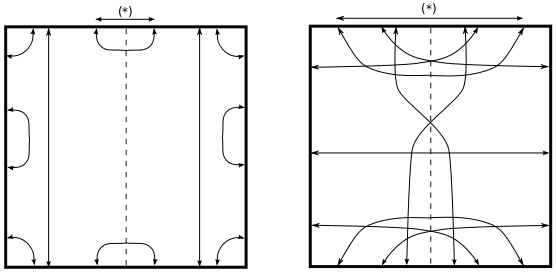}
	\caption{A planar code consisting only of four- and one-qubit stabilizers. We only consider $\sigma_x$-errors and the corresponding plaquette operators here. 
		 As in \figref{fig:checkboard} $\sigma_z$-measurements are performed on the top and bottom boundary. The top and bottom boundary are thus able to store and detect the presence of plaquette anyons. 
		 The logical $Z$ operator is a path of single-qubit $\sigma_z$ operators connecting the top and bottom boundary (c.f.\ the operator $Z'$ in \figref{fig:checkboard}). The dashed line lies midway between the left and right boundary.
		 All error paths in the left part of the figure will not lead to a logical error after error correction, while for the error paths in the right part of the figure a logical $X$ operator is performed.
		 Whether an error path of type (*) is correctable depends not only on its length, but also on its position relative to the boundaries. However, any error path of type (*) and of horizontal length below $L/2$ \emph{is} correctable.}
	\label{fig:paths}
\end{figure}

Adding single-qubit measurements at the boundary spins ensures that in a planar code without three-body interactions all error paths that lead to a logical error after error correction consist of $O(L)$ single-qubit errors. 
\figref{fig:paths} illustrates schematically all error paths of a single anyon that do lead to a logical error after error correction and all that do not. 
Strictly speaking, this figure is only valid for the first anyon, as a concatenation of paths that do not cause an error may lead to an error after error correction.

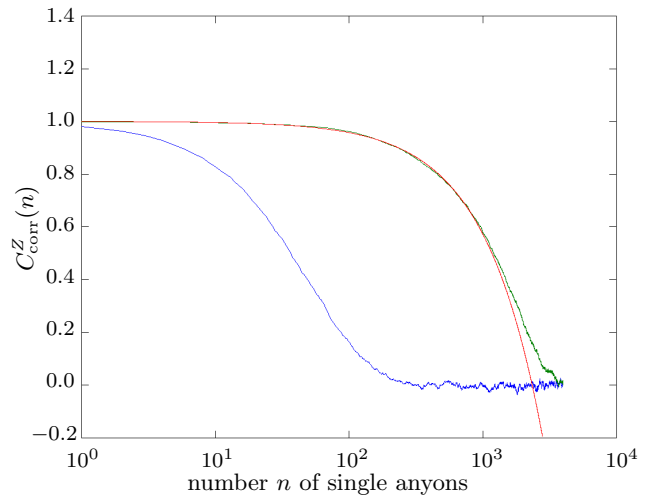
\begin{figure}
  \setlength{\unitlength}{0.5\textwidth}
    \begin{picture}(0.8,0.8)	
	\put(0.04,0.1){\includegraphics[width=0.4\textwidth]{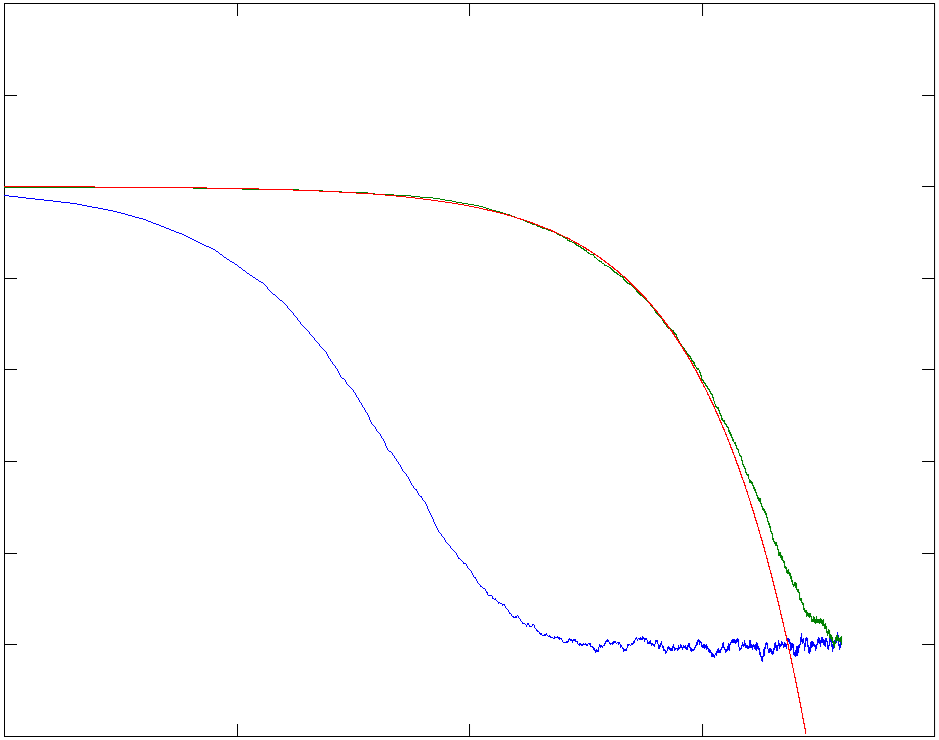}}
	\put(-0.06,0.35){\begin{sideways}$C^Z_{\m{corr}}(n)$\end{sideways}}
	\put(0.2,0.025){number $n$ of single anyons}

	\put(0.02,0.06){\footnotesize $10^0$}
	\put(0.22,0.06){\footnotesize $10^1$}
	\put(0.42,0.06){\footnotesize $10^2$}
	\put(0.62,0.06){\footnotesize $10^3$}
	\put(0.82,0.06){\footnotesize $10^4$}

	\put(-0.032,0.10){\footnotesize $-0.2$}
	\put(-0.01,0.1775){\footnotesize $0.0$}
	\put(-0.01,0.255){\footnotesize $0.2$}
	\put(-0.01,0.3325){\footnotesize $0.4$}
	\put(-0.01,0.41){\footnotesize $0.6$}
	\put(-0.01,0.4875){\footnotesize $0.8$}
	\put(-0.01,0.565){\footnotesize $1.0$}
	\put(-0.01,0.6425){\footnotesize $1.2$}
	\put(-0.01,0.72){\footnotesize $1.4$}

    \end{picture}
  \caption{A planar code of size $L = 64$ consisting only of four- and one-qubit stabilizers in contact with a bath with $\gamma\left((N-1)\Delta\right) / \gamma(0) = 40$, where $N=2L(L+1)$. The vertical axis shows the number $n$ of single anyons that have been created.
		The green curve shows the error-corrected autocorrelation function $C^Z_{\m{corr}}(n)$ and the blue curve the uncorrected autocorrelation function (\eqref{eq:CZcorr} without $\Phi_{\m{corr}}$). Both curves are sampled over $1.2\cdot10^4$ experiments.
 		The red curve shows the analytical prediction \eqref{eq:hcAutocorr} with $c' = 1.12$. 
		We see that performing a single error correction step before the read-out of the stored quantum information allows to enhance its lifetime ($\tau(0.1)$, say) by more than an order of magnitude.}
  \label{fig:hcAutocorr}
\end{figure}

It is in principal possible to find an analytical expression for the probability that the first anyon causes an error in such a planar code of size $L$. 
However, we expect from the discussion of the Gambler's Ruin problem in Appendix~\ref{app:gambler}
that this probability takes the form $\frac{\gamma(0)}{\gamma(0)+\gamma\left((N-1)\Delta\right)}\cdot\frac{2}{L}$, up to some constant factor of order $O(1)$. 
This probability will, due to both factors, be small such that, following the discussion before \eqref{eq:CcorrP}, we expect the error-corrected autocorrelation function after the creation of $n$ single anyons to take the form
\begin{align}\label{eq:hcAutocorr}
 C^Z_{\m{corr}}(n) \simeq 1 - \frac{c'}{L}\cdot \frac{\gamma(0)}{\gamma(0)+\gamma\left((N-1)\Delta\right)}\cdot n\ ,
\end{align}
where $c'$ is a constant.
Of course, this approximation can only be valid as long as $C^Z_{\m{corr}}(n)$ is still relatively close to $1$. 
The numerics in \figref{fig:hcAutocorr} are in excellent agreement with this prediction as long as $C^Z_{\m{corr}}(n) \gtrsim 0.5$.

\begin{figure}
  \setlength{\unitlength}{0.5\textwidth}
    \begin{picture}(0.8,0.8)	
	\put(0.0,0.1){\includegraphics[width=0.4\textwidth]{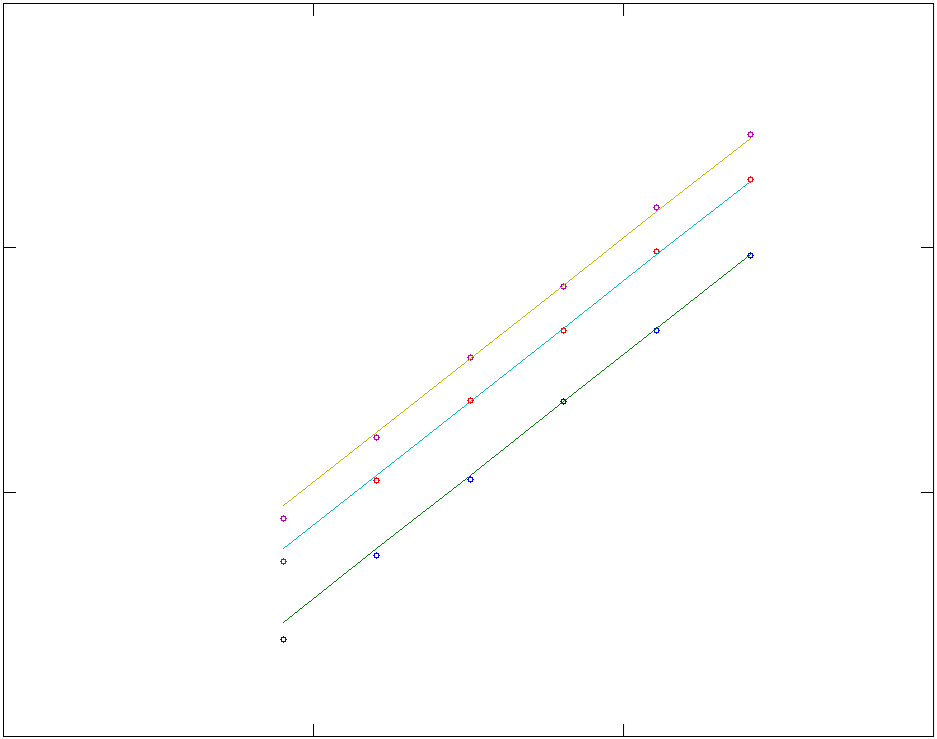}}
	\put(0.35,0.02){size ($L$)}

	\put(-0.02,0.06){\footnotesize $10^0$}
	\put(0.25,0.06){\footnotesize $10^1$}
	\put(0.52,0.06){\footnotesize $10^2$}
	\put(0.79,0.06){\footnotesize $10^3$}

	\put(-0.055,0.10){\footnotesize $10^1$}
	\put(-0.055,0.306){\footnotesize $10^2$}
	\put(-0.055,0.512){\footnotesize $10^3$}
	\put(-0.055,0.718){\footnotesize $10^4$}

    \end{picture}
  \caption{The number of single anyons created during the lifetime $\tau(\eps)$ of the memory as a function of the lattice size $L$. The points show numerical results obtained for a bath with $\gamma\left((N-1)\Delta\right) / \gamma(0) = 40$, where $N=2L(L+1)$,
		  the lines show the analytical prediction \eqref{eq:nAnyons} with $c' = 1.12$. From bottom to top we have $\eps = 0.1 , 0.2 , 0.3$.}
  \label{fig:nAnyons}
\end{figure}
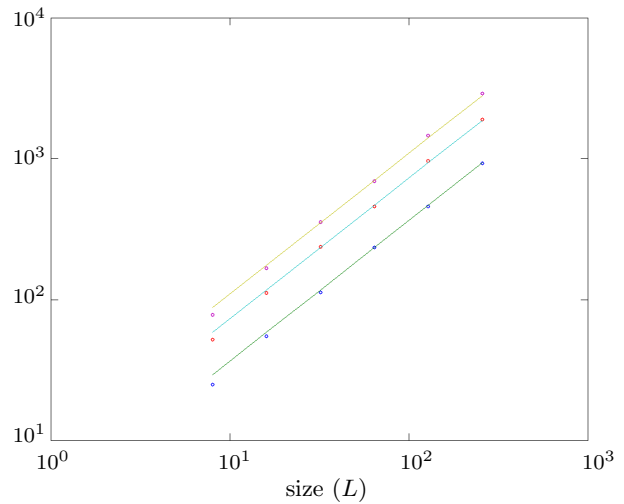

Correspondingly, the number of single anyons that are created during the lifetime $\tau(\eps)$ of the stored quantum information will be 
\begin{align}\label{eq:nAnyons}
 \frac{\eps \cdot L}{c'}\cdot \frac{\gamma(0)+\gamma\left((N-1)\Delta\right)}{\gamma(0)}\ .
\end{align}
This prediction is compared with numerical simulations in \figref{fig:nAnyons}.
Since there are for each kind of anyon $2\cdot(2L+1)$ qubits where a single anyon can be created (including the ones on which single-qubit measurements are performed), we obtain a lifetime
\begin{align}\label{eq:hcLifetime}
 \tau(\eps) &\simeq \frac{\eps \cdot L}{c'} \cdot \frac{\gamma(0)+\gamma\left((N-1)\Delta\right)}{\gamma(0)} \nn\\&\quad \cdot\left[2\cdot(2L+1)\cdot \gamma\left(-(N-1)\Delta\right)\right]^{-1} \nn\\
  &\simeq \frac{\eps}{4c'} \cdot \frac{\exp\left((N-1)\beta\Delta\right)}{\gamma(0)}\ .
\end{align}
In the last step, we applied detailed balance and $\gamma(0) \ll \gamma\left((N-1)\Delta\right)$. 
Since $N=2L(L+1)$, the lifetime of the memory grows exponentially in $L^2$. This provides, to the best of our knowledge, the best scaling of a quantum memory lifetime with the linear size found so far in at most three dimensions.
However, we have noted that strictly speaking $L$ is bounded through the breakdown of the perturbative treatment of the underlying honeycomb Hamiltonian. So similarly as in \cite{Haah11c} the found scaling of the lifetime is only valid up to some optimal $L^*$.
Still, our optimal $L^*$ is generic in the sense that changing it by a small integer will not have a drastic effect on the lifetime.

We note that the only time-scale that entered \eqref{eq:hcLifetime} was the time needed to create a single anyon, while $\gamma(0)$ only entered through a probability. 
The former canceled by detailed balance such that the only remaining time-scale was the hopping time $1/\gamma(0)$. 
In the toric case, anyons can only be created pairwise, such that the gap above the anyonic vacuum is increased from $(N-1)\Delta$ (as in the planar case) to $2(N-2)\Delta$, leading to a lifetime increasing as $\sim\exp(2(N-2)\beta\Delta)$, in contrast to \eqref{eq:hcLifetime}. 
Besides complicating error correction, the realistic case of open boundaries thus also reduces the exponentially increasing factor in the lifetime of the memory to (almost) its square root.

\subsection{Topological order at finite temperature}

The above results apply to the code during its thermalization. However, it is also interesting to study its properties once it reaches thermal equilibrium. 
Clearly the memory will have completely decohered by this point, since the probability of the system being in any of its ground states will be equal. 
However, we can assess whether topological order is present. It is known that, for the non-interacting case of \eqref{eq:Kitaev}, the thermal state is not topologically ordered for any finite temperature. 
However, here we show that the interactions of \eqref{eq:holeRepulsion} allow the topological order to remain stable for all finite temperatures at which the perturbative derivation of the Hamiltonian \eqref{eq:holeRepulsion} is valid.

To determine whether topological order is present we can use one of the topological order parameters designed for mixed states, such as the anyonic topological entropy \cite{Wootton12}. 
This requires the plaquettes and vertices of the code to be split into three regions, $A$, $B$, and $C$. 
These can be defined arbitrarily, except that $A$ and $B$ must be bounded, $B$ must enclose $A$, $C$ must enclose $B$ and the number of plaquettes and vertices in each region must be $O(L^2)$. 
The value of the entropy depends on how well the anyon configuration within $B$ can be used to deduce the net anyonic occupation of $A$. 
If this can be done perfectly, the entropy takes its maximum possible value (for the planar code) of $2 \ln 2$, signaling that the state is topologically ordered.

Let us consider the thermal state of the Hamiltonian \eqref{eq:holeRepulsion}. 
The Hamiltonian is symmetric under exchange of anyons and anyon holes. 
Since the gap above the ground states with all anyons or all holes grows as $L^2$, the thermal state in the thermodynamic limit will be an equally weighted mixture of the all hole or all anyon states. 
As such, measurement of the occupancy of any plaquette can be used to determine the occupancies of all plaquettes. The region $B$ used in the definition of the anyonic topological entropy can then clearly deduce the net occupancy of the region $A$. 
The value for this entropy therefore takes its maximum value of $2 \ln 2$, and the state is found to be topologically ordered.

\section{Conclusions}

Stabilizer Hamiltonians with local interactions in 2D do not, and in 3D seem not to, offer the possibility to passively store quantum states for a time that can be made arbitrarily larger than the relevant microscopic time-scales.
Inducing long-range interactions between the excitations of a 2D Hamiltonian whose ground states are topologically ordered seems thus the most promising approach towards a realistic proposal for a quantum memory.
For such long-range interactions, the influence of the boundary of the memory (which every realistic memory will have) is not negligible even for large $L$. 
We discussed two Hamiltonians proposed in the recent literature and showed that for those the boundary becomes in fact dominant.
Operationally, this fact becomes relevant during the error correction step before the read-out of the stored quantum state. 
We showed that the classical algorithm that determines how to best remove the anyonic defects has to depend on the error model, the memory Hamiltonian and the boundary conditions.

With long-range repulsion between the anyons, the energy to add a further anyon increases with the number of already existing anyons, leading to a vanishing anyon production rate.
If the production rates approach their equilibrium values, the production of unpaired anyons on the boundaries will outweigh the production of anyon pairs in the bulk.
The influence of the boundary anyons is beneficial since they lead to an effective bias of the bulk anyons to move towards the closer boundary.

We discussed how a planar code with topological protection of the stored qubit can be obtained from a honeycomb model with two-qubit Ising coupling and the ability to perform single-qubit measurements on boundary qubits.
In a resonant regime of a coupling of the honeycomb to cavity modes a very strong suppression of the anyon creation rate is obtained. Furthermore, most created anyons will immediately be reabsorbed by their creating boundary.
In conclusion, we have found in this regime a lifetime that grows exponentially in $L^2$, allowing in principle to reach macroscopic storage times even at moderately high temperatures.
The non-local anyon interactions in the obtained effective Hamiltonian are so strong, that the system is topologically ordered at any finite temperature for large enough $L$, 
as long as the perturbatively derived Hamiltonian \eqref{eq:holeRepulsion} describes the dynamics of the memory accurately.

\section{Acknowledgments}
We acknowledge useful discussions with Fabio Pedrocchi. This work was supported by the Swiss NSF, NCCR Nanoscience, and NCCR QSIT.

\newpage
\appendix

\section{Determining the weight of the edges when anyons perform a random walk}\label{app:diffusion}

Given an anyon at time $t$ at position $\vec{a}=(a_1 , a_2)$, where $a_1$ is the distance from the left boundary and $a_2$ the distance from the upper boundary, what is the probability $\Pr\left[\vec{a},t',t\right]$ that it has been created at a time $0 < t' < t$ at the upper boundary?
The answer can be obtained by summing \eqref{eq:diffusion} over all spins $0, \ldots, L$ on the upper boundary, so
\begin{align}\label{eq:diffusionProb}
 &\Pr\left[\vec{a},t',t\right] \nn\\
&\simeq \int_0^{L+1} \m{d}x\, \frac{1}{4\pi D(t-t')} \cdot e^{-(a_2^2+(a_1-x)^2)/4D(t-t')} \nn\\
& = \frac{1}{4}\frac{1}{\sqrt{\pi D(t-t')}} \cdot e^{-a_2^2/4D(t-t')}  \nn\\&\quad \cdot \left\lbrace\m{Erf}\left[\frac{a_1}{2\sqrt{D(t-t')}}\right]+\m{Erf}\left[\frac{L+1-a_1}{2\sqrt{D(t-t')}}\right]\right\rbrace\ .
\end{align}
The total probability that that an anyon that is at time $t$ at position $\vec{a}$ has been created on the upper boundary is then
\begin{align}
 \Pr\left[\vec{a},t\right] = \int_0^t \m{d}t'\, \Pr\left[\vec{a},t',t\right] \cdot \gamma_{\m{boundary}}(t') \cdot \xi(t',t)\ ,
\end{align}
where $\gamma_{\m{boundary}}(t')$ is the creation rate of anyons on the boundary at time $t'$ and $\xi(t',t)$ is the probability that an anyon that has been created at a boundary at time $t'$ does still exist at time $t$.
Unfortunately, the time integration cannot be performed in closed form even if we take $\gamma_{\m{boundary}}$ and $\xi$ to be constant and $\left\lbrace \ldots \right\rbrace = 2$ in \eqref{eq:diffusionProb} (corresponding to the $L\ra\infty$ limit).

Still, the above analytics is enough to find reasonable weights for the edges in our error correction algorithm. 
First, let us note that we can find a similar expression like the one above for the situation where we have anyons at postion $\vec{a}$ and $\vec{b}$ and are interested in the probability that they have been jointly created in the bulk. 
The exponential factor becomes in this case $\sim\exp\left[-(\vec{a}-\vec{b})^2/8D(t-t')\right]$. Intuitively, the distance vector of the two anyons performs a diffusive motion with diffusion constant $2D$.
A sensible choice for the weight of an edge between two anyons is therefore the square of their Euclidean distance, while for an edge connecting an anyon to its closer boundary we take the weight to be \emph{twice} the square of the distance.
All terms depending on time, the position of the anyons relative to the boundaries and the creation rates give then additive logarithmic correction terms to these weights.

Note that in the matching which is obtained at the end of the algortihm the number of edges connecting two real anyons is always equal to the number of edges connecting two virtual ones.
Furthermore, adding the same term to the weights of all edges or multiplying all weights with the same positive term does not change the result of the algorithm. 
We may therefore give the edges that connect two virtual anyons a weight which takes all logarithmic correction terms into account. 

Formally, let $d^2$ denote the square of the Euclidean distance of the pair we are interested in. 
Then, giving edges that connect two real anyons a weight $\alpha\cdot d^2 + \beta_2$ (with $\alpha > 0$), edges that connect a real anyon with a virtual one a weight $\alpha\cdot 2d^2 + \beta_1$ and edges that connect two virtual anyons a weight $0$ 
is equivalent to giving them weights $d^2 + (\beta_2 - \beta_1) /  \alpha$, $2d^2$ and $-\beta_1/\alpha$ respectively,
which is again equivalent to giving them weights $d^2$, $2d^2$ and $(\beta_2 - 2\beta_1) /  \alpha$ respectively.
Rather than calculating the term $(\beta_2 - 2\beta_1) /  \alpha$ analytically (which we cannot) we may then consider the weight of the virtual edges as a single optimization parameter of our algorithm. 
We have found numerically that in the systems we are interested in varying the weight of the purely virtual edges offers hardly room for improvement of the memory lifetime. As before, we will thus take their weight to be zero throughout Sec.~\ref{sec:longrange}.
However, there are certainly regimes where these weights become relevant. If, for example, the rate for creation of anyon pairs in the bulk is vanishing against the rate for creation on the boundaries, the edges between virtual anyons should be given a large weight.

\section{Bounds on the self-consistent anyon density}\label{app:estimates}

Using \eqref{eq:eeq} to eliminate $e_{\m{eq}}$ in \eqref{eq:neq} we find
\begin{align}
 1 = n_{\m{eq}} \cdot \left[\exp\left(\beta (\Delta + A (L^2 n_{\m{eq}} - 1))\right) + 1\right] := f(n_{\m{eq}})\ .
\end{align}
Since $f(x)$ is monotonically increasing in $x$ we have that $x < n_{\m{eq}} \Leftrightarrow f(x) < 1$.
We have
\begin{align}
 f(\frac{1}{L^2}) = \frac{1}{L^2}\left[\exp\left(\beta\Delta\right)+1\right]
\end{align}
and
\begin{align}
 f(\frac{1}{L^{2-\eps}}) > \frac{1}{L^{2-\eps}}\cdot\exp\left(\beta A (L^\eps-1)\right)\ ,
\end{align}
yielding the estimates \eqref{eq:bound1} and \eqref{eq:bound2}.

\section{The Gambler's Ruin}\label{app:gambler}

In the usual version of the Gambler's Ruin problem, the probability of winning or losing is the same for every amount of money the gambler possesses.
In the anyonic hopping problem we are interested in, the probability of moving towards the boundary is not excactly independent of the distance to the boundary.
Since the rate for absorbtion of an anyon by a boundary $\gamma(\m{absorb})$ is usually much higher than the hopping rate $\gamma(0)$, the anyon is biased to move towards the boundary if it is at a position next to it.
This is in particular the case immediately after the anyon has been created.

We therefore study the following problem. Consider a lattice with rows $0, \ldots, L-1$ and an anyon initially in row $0$ (it has just been created). 
We are only interested in hoppings between the rows and ignore all movements within the same row. 
In row $0$ there is a probability $\tilde{p} = \frac{\gamma(\m{absorb})}{\gamma(\m{absorb})+\gamma(0)}$ that the anyon is absorbed by the adjacent boundary and a probability $1-\tilde{p}$ that it hops to row $1$ (similarly in row $L-1$, but this will be irrelevant here).
In all other rows, the anyon hops to both adjacent rows with equal probability.
What is the probability that the anyon reaches row $L/2$ (let $L$ be even for simplicity) at least once before it is reabsorbed?

Let $p_i$ the denote the probability to reach row $L/2$ from row $i$. The boundary conditions are 
\begin{align}\label{eq:bdry}
 p_0 = (1-\tilde{p}) \cdot p_1
\end{align}
and 
\begin{align}
 p_{L/2} = 1\ .
\end{align}
For $1 \leq i \leq L/2-1$ we have $p_i = \half(p_{i-1}+p_{i+1})$, or equivalently $p_{i+1} - p_i = p_i - p_{i-1}$.
We find
\begin{align}
 1 = p_{L/2} &= \sum_{i=1}^{L/2-1}(p_{i+1} - p_i)+p_1 \nn\\&= (L/2-1)\cdot(p_1-p_0)+p_1
\end{align}
Using \eqref{eq:bdry} to eliminate $p_1$ we arrive at
\begin{align}
 p_0 = \frac{1-\tilde{p}}{(L/2-1)\cdot\tilde{p}+1}\ ,
\end{align}
which for $1-\tilde{p} \ll 1$ (i.e.\ $\gamma(0) \ll \gamma(\m{absorb})$) simplifies to
\begin{align}
 p_0 \simeq (1-\tilde{p})\cdot\frac{2}{L} \simeq \frac{\gamma(0)}{\gamma(\m{absorb})}\cdot\frac{2}{L}\ .
\end{align}

Now let $n_i$ denote the expected number of hoppings (perpendicular to the boundary under interest) necessary to reach row $L/2$, assuming that it is eventually reached.
The boundary conditions are
\begin{align}
 n_0 = 1 + n_1
\end{align}
and
\begin{align}
 n_{L/2} = 0\ .
\end{align}
For $1 \leq i \leq L/2-1$ we have the recursion $n_i = 1 + \half(n_{i-1}+n_{i+1})$ or equivalently $n_{i+1} - n_i = n_i - n_{i-1} - 2$. 
From this we expect a quadratic expression for $n_i$ and thus make the \emph{Ansatz}
\begin{align}
 n_i = a\cdot i^2 + b\cdot i + c
\end{align}
with the unique solution $a=-1$, $b=0$ and $c=(L/2)^2$. We conclude that $n_0 = (L/2)^2$.


\begin{thebibliography}{10}
\bibitem{eliminates_comment}
Let the `height' $h$ be given by the number of plaquette operators from top to
  bottom and the `width' $w$ by the number of star operators from left to
  right. The total number of qubits is then given by $2hw+h+w+1$, the number of
  plaquette operators by $h(w+1)$ and the number of star operators by $(h+1)w$.
  Since each stabilizer condition eliminates half of all $2^{2hw+h+w+1}$
  degrees of freedom, we are left with $2$ of them.
\bibitem{super_comment}
Note that for a super-Ohmic bath we have $\gamma (0)=0$, forbidding the direct
  hopping of anyons and heavily suppressing their diffusion. In this case, only
  `indirect hopping' \cite {Chesi10} is possible, in which a new pair of anyons
  is created next to an existing one and the existing one fuses with one of
  them, leading to an effective movement of the already existing anyon.
\bibitem{3qubit_comment}
We assume in this section that the energy penalty for both three- and four-qubit stabilizer violations have the same strength $\Delta$. This can be achieved by appropriately desigend perturbative gadgets \cite{Kempe06}. 
However, we will see that the stability of the memory for large $L$ is due to the second term in Eq. (17), i.e.\ the inter-anyonic interaction, and that the bare anyon energy cost $\Delta$ becomes in fact irrelevant. 
In contrast, in Section IV, we will assume that the stabilizer operators are obtained from the honeycomb model such that there is no energy penalty associated with three-qubit stabilizer operators.
\end{thebibliography}
\end{document}